\documentclass[aps,prl,preprint,floatfix]{revtex4-1}
\usepackage{amsmath}
\usepackage{amsfonts}
\usepackage{amssymb}
\usepackage{bm}
\def\*#1{\mathbf{#1}}
\usepackage[usenames,dvipsnames]{color} 
\usepackage{ulem}
\usepackage{graphicx}
\graphicspath{ {./figs/} }
\usepackage{epstopdf}

\usepackage{csquotes}
\usepackage[sort&compress]{natbib}

\usepackage [english]{babel}
\usepackage[utf8]{inputenc}
\usepackage[T1]{fontenc}

\usepackage{gensymb}

\usepackage{hyperref}
\hypersetup{
  colorlinks   = true, 
  urlcolor     = black, 
  linkcolor    = red, 
  citecolor   = blue 
}

\usepackage{tabularx}
\newcolumntype{Y}{>{\centering\arraybackslash}X}  
\DeclareUnicodeCharacter{2212}{-}

\begin{document}

\title{
Origin of the non-linear elastic behavior of silicate glasses
}

\author{Zhen Zhang}
\affiliation{ Center for Alloy Innovation and Design, 
State Key Laboratory for Mechanical Behavior of Materials, 
Xi’an Jiaotong University, Xi’an 710049, China}
\affiliation{Laboratoire Charles Coulomb (L2C), 
University of Montpellier and CNRS, F-34095 Montpellier, France}

\author{Simona Ispas}
\affiliation{Laboratoire Charles Coulomb (L2C), 
University of Montpellier and CNRS, F-34095 Montpellier, France}

\author{Walter Kob}
\email[Corresponding author: ]{walter.kob@umontpellier.fr}
\affiliation{Laboratoire Charles Coulomb (L2C),
University of Montpellier and CNRS, F-34095 Montpellier, France}
\date{\today}

\begin{abstract}  
For small tension the response of a solid to an applied stress is given by Hooke's law. Outside this linear regime the relation between stress and strain is no longer universal and at present there is no satisfactory insight on how to connect for disordered materials the stress-strain relation to the microscopic properties of the system. Here we use atomistic computer simulations to establish this connection for the case of silicate glasses containing modifiers. By probing how in the highly non-linear regime the stress-strain curve depends on composition, we are able to identify the microscopic mechanisms that are responsible for the complex dependence of stress on strain, notably the presence of an unexpected quasi-plateau in the tangent modulus.
We trace back this dependence to the mobility of the modifiers which, without leaving their cage or modifying the topology of the network, are able to relieve the local stresses. Since the identified mechanism is general, the results obtained in this study will also be helpful for understanding the mechanical response of other disordered materials. \\

\end{abstract}

\maketitle

\bigskip

In contrast to crystalline solids, disordered materials such as oxide glasses, gels etc.~have the advantage that their composition can be chosen basically at will, thus allowing for a great flexibility for tuning their mechanical, optical, and electric properties~\cite{binder_glassy_2011,varshneya2019fundamentals,rubinstein2003polymer}. This feature, in combination with the structural isotropy of the material, makes them highly attractive for many applications such as windows panes, optical fibers, polymeric materials, high strength device, food, micro-devices, etc. The modification of material properties by a change of composition is, however, not always a trivial task since often the relation between composition and property is highly non-linear, as it is, e.g., the case in polymeric systems to which one adds plasicizers to enhance the flexibility of the  material~\cite{rubinstein2003polymer}. This strong dependence is related to the fact that the structural and dynamical properties of disordered systems are influenced by a multitude of competing and counteracting mechanisms, many of which have not yet been understood.

Silica glasses, arguably the most important class of oxide glasses, are known to have a complex non-linear strain-dependence of the elastic modulus~\cite{mallinder1964elastic,krause_deviations_1979,kurkjian1985strength,gupta_intrinsic_2005}, which has been shown to be important for questions relating to strength~\cite{gupta_intrinsic_2005,kurkjian2010strength,wondraczek_towards_2011,mihai2017nonlinear}. Understanding this non-linear behavior is thus of fundamental interest for glass physics and also of practical importance for the application of these materials, notably in applications under extreme conditions~\cite{suhir1992elastic,wallenberger2010commercial,wiederhorn_fracture_1974}. Due to its importance, silica glass has been studied extensively and it has been found that its tangent modulus $E_{\rm t}$ increases with strain up to a maximum at around $\varepsilon=0.1$ and then decreases again~\cite{gupta_intrinsic_2005,pedone_molecular_2008,yuan_molecular_2012}, a behavior that has been termed ``anomalous'' since most glasses show a decrease of $E_{\rm t}$ with strain. The maximum of $E_{\rm t}$ upon tension has been associated with the observation of a maximum in the isothermal compressibility on hydrostatic compression
~\cite{pedone_molecular_2008,bridgman_compressibility_1925,bridgman_high_1938,meade1987frequency,tsiok1998logarithmic},  the origin of which has been attributed to the presence of floppy modes in a mixture of high- and low-density amorphous phases in a recent experimental study~\cite{clark_mechanisms_2014}. 

The elastic behavior of modifier-containing glasses, such as alkali silicate, has been understood much less. It is found that the anomalous properties of silica glass disappear with the addition of modifiers, e.g. Na,~\cite{wiederhorn_fracture_1974,yuan_molecular_2012}. The transition from anomalous to normal has been attributed to the reduced three-dimension connectivity due to the presence of the network modifiers, although the behavior of the modifiers during deformation has so far not been understood. We also note that previous studies  mainly focused on the elastic properties of the glasses at small strains (typically <5\%)~\cite{gupta_intrinsic_2005}, whereas the elastic behavior at higher strains (important for applications at extreme conditions) is basically unknown. The goal of the present work is therefore to probe the highly non-linear regime of the elastic properties of silicate glasses and to identify the relevant mechanisms that give rise to the complex strain and composition dependence of the stress-strain curves of these systems. As we will see, these mechanisms are very general and hence the obtained insights are expected to be useful to understand also the mechanical properties of other disordered systems.

The compositions we investigate are pure SiO$_2$, the binary mixture Na$_2$O-$x$SiO$_2$ (NS$x$) with $x=3, 5, 10$, and the mixtures A$_2$O-3SiO$_2$ (AS3) with A=Li, Na, K.
 These systems are of great importance in various fields, such as the glass-making industries and geosciences. The molecular dynamics simulations were performed using an effective potential~(SHIK)~\cite{sundararaman_new_2018,sundararaman_new_2019}
which has been shown to give a reliable description of the
structural and mechanical properties of sodium silicate glasses~\cite{zhang_potential_2020}. More details regarding the simulations are given in the Methods.
We emphasize that the simulation parameters (sample size, cooling rate, strain rate, etc.) were chosen such that the presented results are robust~\cite{zhang_potential_2020,zhang_thesis_2020}.

\begin{figure}[ht]
\includegraphics[width=0.95\columnwidth]{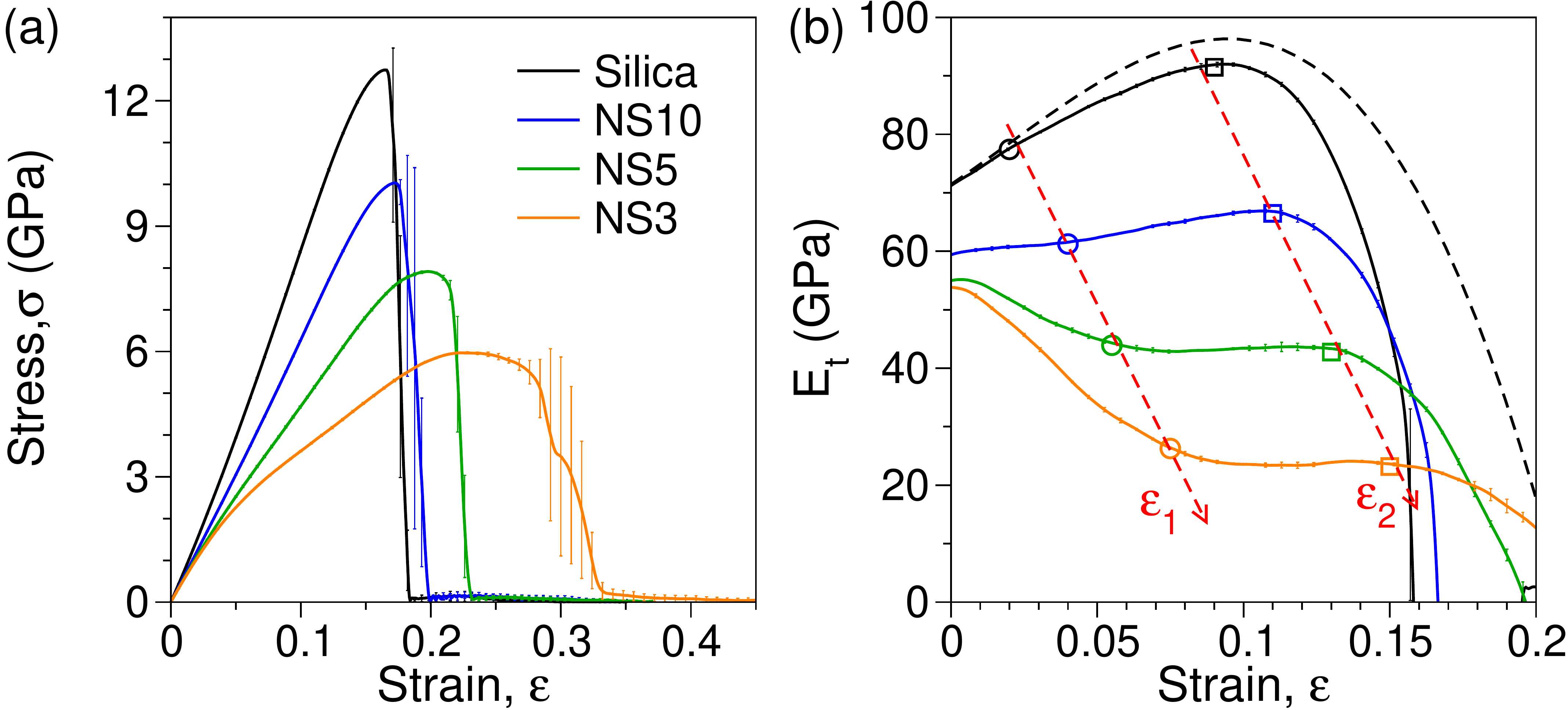}
\includegraphics[width=0.95\columnwidth]{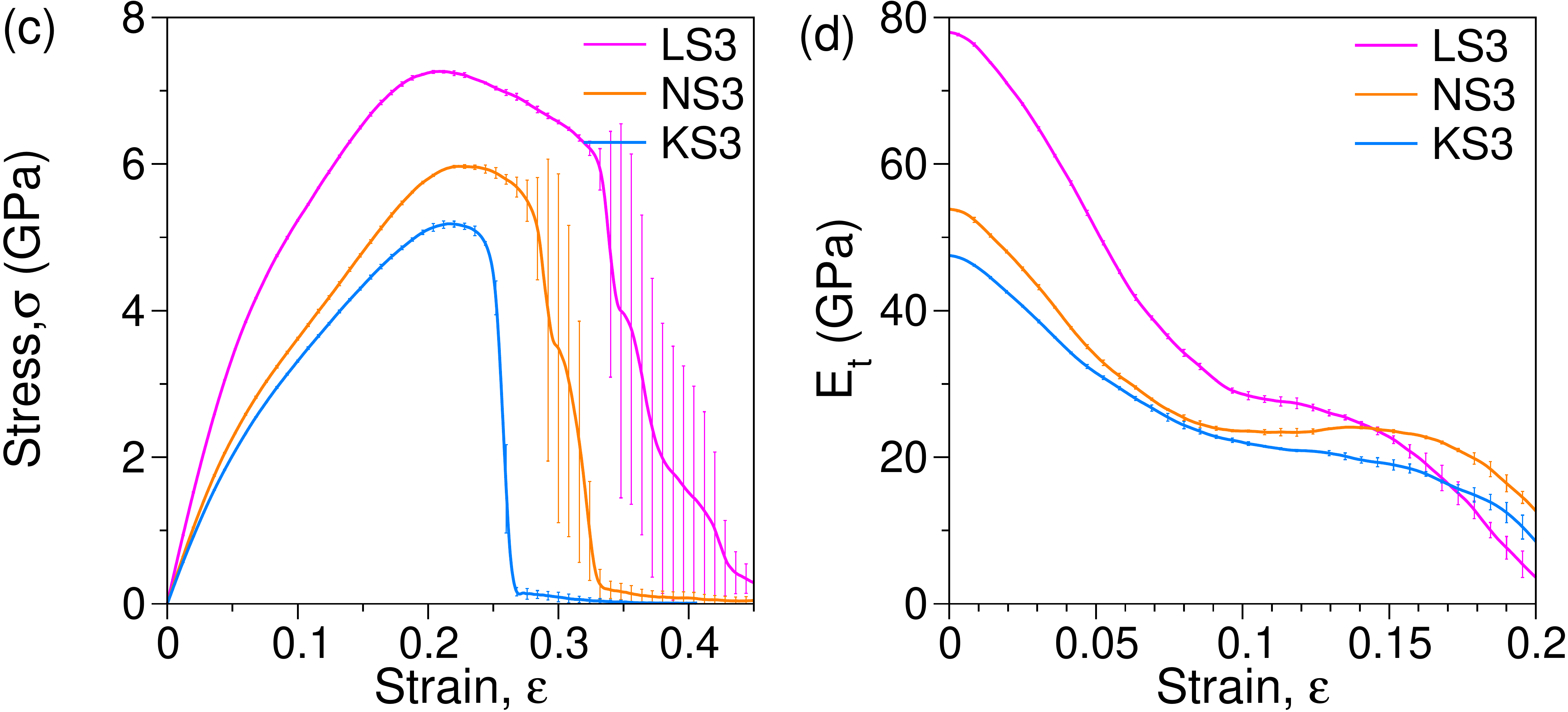}
\caption{(a) and (b) are, respectively, the stress-strain curves and the corresponding tangent modulus $E_t$ for the NSx glasses. 
The black dashed curve is the fit to experimental data of silica glass fibers up to 7\% strain~\cite{guerette_nonlinear_2016}. The red dashed lines are guide to the eye, highlighting two characteristic strains, $\varepsilon_1$ and $\varepsilon_2$, which are indicated by the symbols, see text for details. (c) and (d) are, respectively, the stress-strain curves and the corresponding $E_t$ for the AS3 glasses.  
All three compositions show a bend in their tangent modulus at $\varepsilon\approx0.08$ but this feature is most pronounced for NS3.
}
\label{fig1_AS-ss-E-bulk}
\end{figure}

Figure~\ref{fig1_AS-ss-E-bulk} presents the macroscopic stress-strain (SS) curves and their derivatives, i.e., the tangent modulus $E_t$ of the glasses. (In Fig.~\ref{SI_fig_failure-Y-compare} we show that the failure strain, strength and Young's modulus of the simulated glasses are in good agreement with experimental measurements, demonstrating the reliability of our simulations.) From panel (a) one recognizes that the addition of Na$_2$O to the SiO$_2$ glass leads to a decrease of the strength, i.e.,~the maximum attainable stress before failure, and an increase of the ductility, i.e., the ability of a material to elongate under tensile loading.
Also, the decreasing slope of the SS curve with the addition of Na shows that the presence of the network modifiers softens the network at large strains. 

In panel (b) this influence by the modifier is quantified and one recognizes that already at $\epsilon=0$ the addition of 25\% Na$_2$O reduces the Young's modulus of the glass from around 71 GPa for silica to $56$ GPa for NS3, in good agreement with experimental findings~\cite{bansal_handbook_1986,lower_inert_2004,januchta_elasticity_2019}.  This softening of the glass is directly related to the decrease in network connectivity as the concentration of modifiers increases~\cite{rouxel_elastic_2007}. From this graph one also recognizes that the non-linear elastic behavior of the glasses can be divided into three regimes, each of which has a different strain dependence of $E_t$, marked by the two arrows which indicate the two  strains $\varepsilon_1$ and $\varepsilon_2$ that delimit these regions. (Note that there is no explicit mathematical definition of $\epsilon_1$ and $\epsilon_2$ but instead they are just rough estimates of the regions' boundaries.) For $\varepsilon<\varepsilon_1$ the addition of Na makes that $E_t(\varepsilon)$ gradually transform from anomalous (i.e.~$E_t$ increases) to intermediate ($E_t$ remains almost constant) to normal ($E_t$ decreases), in agreement with experimental findings~\cite{gupta_intrinsic_2005,krause_deviations_1979}. In the second regime, i.e., $\varepsilon_1<\varepsilon<\varepsilon_2$, one finds for the Na-rich glasses, NS5 and NS3, that  $E_t$ is basically independent of strain. This result is surprising since naively one anticipates a continuous softening of the glass upon increasing strain. In order to ensure that this finding is not just a particularity of the interaction potential, we have also calculated $E_t$ using other popular potentials (although computationally more expensive) and the results (see Fig.~\ref{SI_fig_ns3-bend-mix-pot}) show that the observed plateau in $E_t$ for the Na-rich glasses is a robust feature. If the Na concentration is lowered, the $\varepsilon$-dependence of $E_t$ in this second regime becomes also anomalous, although the slope of $E_t$ is different from the one at small strains, from which we conclude that  the mentioned plateau is related to the high concentration of network modifiers. For the third regime, $\varepsilon > \varepsilon_2$, all the $E_t$ curves show a rapid decrease which can be attributed to global yielding of the glasses.     

Figures~\ref{fig1_AS-ss-E-bulk}(c) and (d) show the SS and the resulting $E_t$ curves for the different AS3 glasses. From panel (c) one recognizes that the glasses with small alkali atoms have not only higher strength but they are also more ductile. Thus reducing the size of the modifiers is a viable strategy to improve the toughness of the glass, in accordance with experimental findings~\cite{kurkjian_intrinsic_2001}. More important for the subject of the present study is the fact that also LS3 and KS3 exhibit a similar $\varepsilon$-dependence of $E_t$ as observed for NS3, notably the plateau at intermediate strains, panel (d). This indicates that the unexpected non-linear $\varepsilon$-dependence of $E_t$ is universal for alkali-containing glasses, hinting a common origin and in the following we will identify the microscopic mechanism leading to this behavior.

\begin{figure}[ht]
\includegraphics[width=0.55\columnwidth]{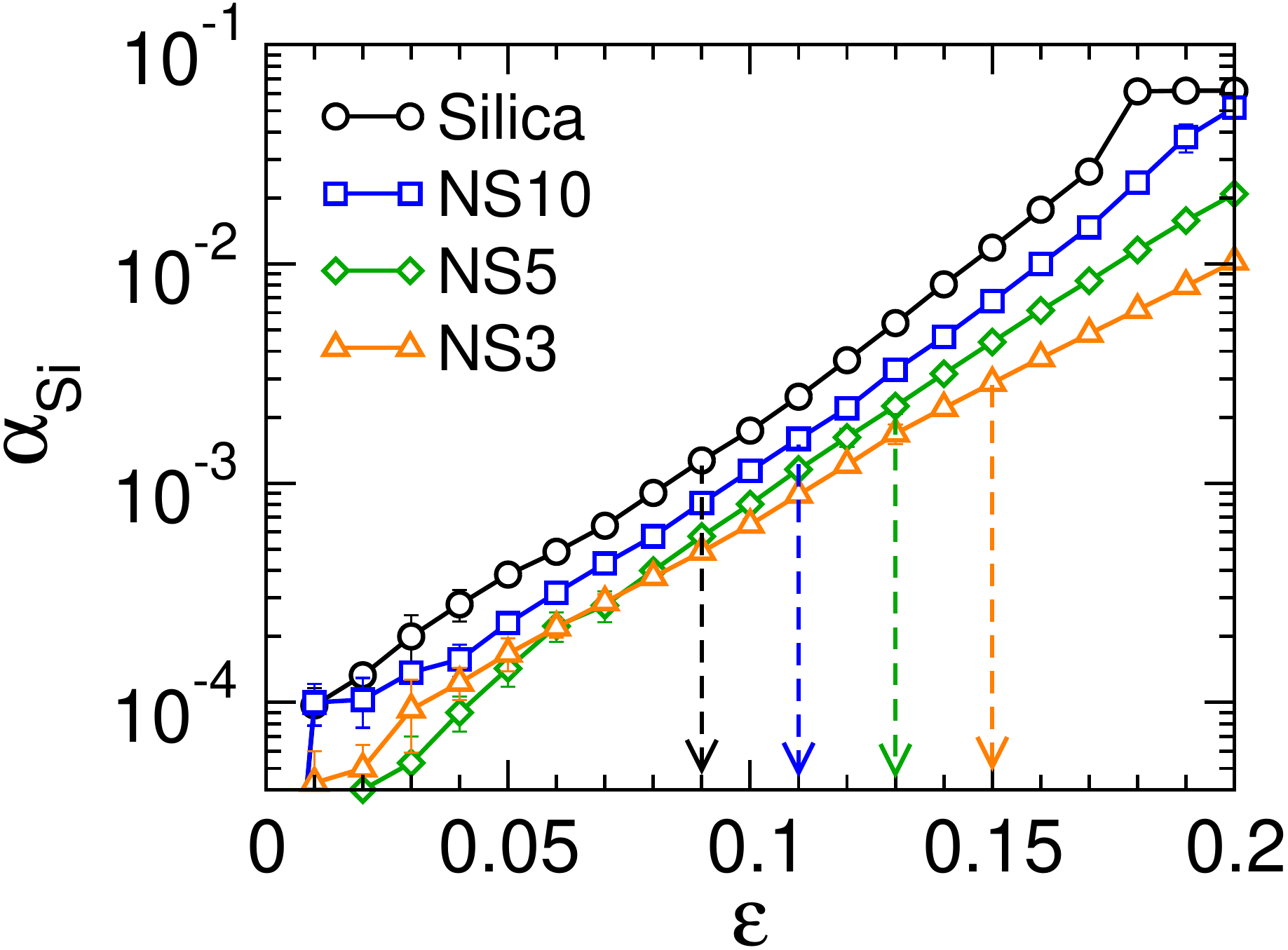}
\caption{
Fraction of Si atoms that have changed their bonding environment $\alpha_{\rm Si}$, see main text for definition. The vertical arrows from left to right indicate approximately the $\epsilon_2$ of silica, NS10, NS5, and NS3, respectively.
}
\label{fig2_nc-si-bonding}
\end{figure}

Usually the network topology (NT) is a convenient structural observable for understanding the mechanical behavior of glasses~\cite{mauro2011topological,bauchy2019deciphering,zheng_intjouapplglasssci_11_432_2020} and therefore we investigate its dependence on strain.  Here we characterize the change of the NT by mean of $\alpha_{\rm Si}(\epsilon)$, the fraction of Si atoms that have changed their bonding partners if the strain is increased from zero to $\varepsilon$: $\alpha_{\rm Si}$=$N'_{\rm Si}$/$N_{\rm Si}$, where $N'_{\rm Si}$ is the number of Si atoms which have at least one of their neighbors changed with respect to the configuration at $\epsilon=0$ and $N_{\rm Si}$ is the total number of Si atoms.
Figure \ref{fig2_nc-si-bonding} shows that $\alpha_{\rm Si}$ is small for all relevant strains, demonstrating that the bonding environment of all network atoms is basically unchanged ($\alpha_{\rm Si}$ <0.3\%  up to $\varepsilon_2$). In other words, the Si-O network remains basically intact during tension up to intermediate strains, a result that is also found for the glasses with Li or K, see Fig.~\ref{SI_fig_nc-si-bonding-as3}. Hence we conclude that the decrease of $E_t$ with strain is not due to a changing topology of the SiO network. \\

\begin{figure}[ht]
\includegraphics[width=0.95\columnwidth]{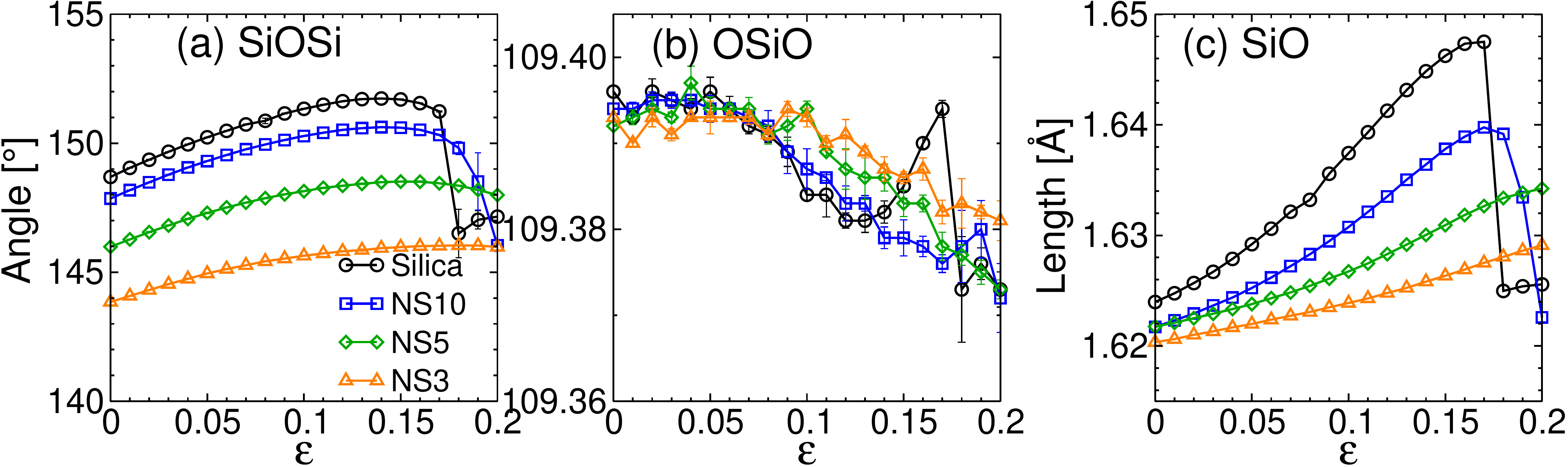}
\includegraphics[width=0.95\columnwidth]{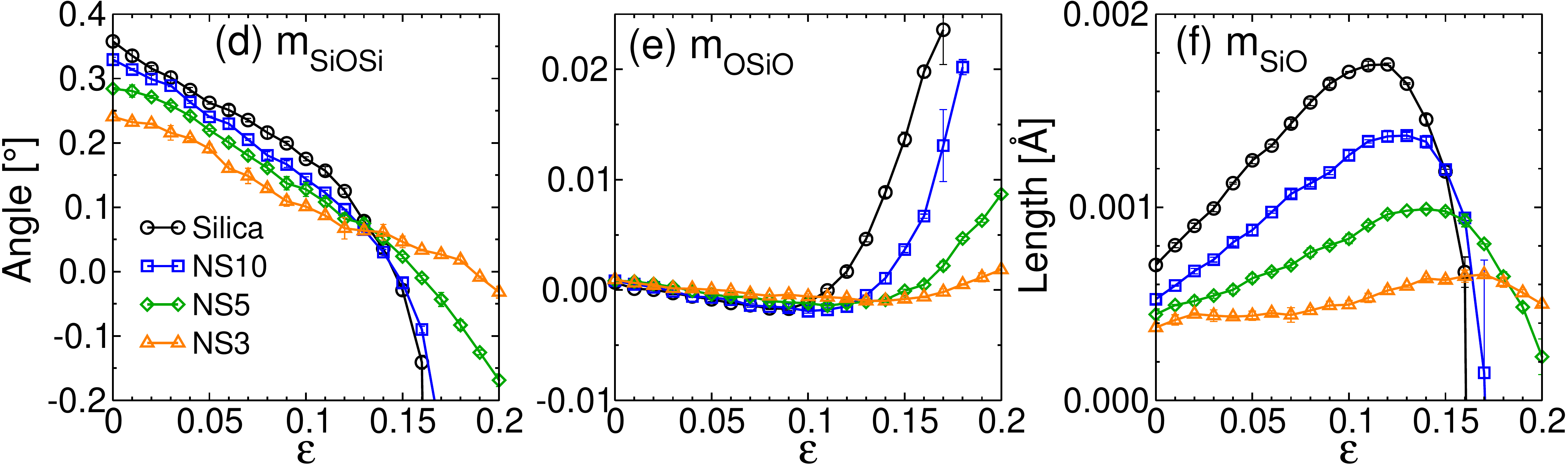}
\caption{ Structural quantities that characterize various deformation modes of the Si-O network. (a)-(c): The mean values. (d)-(f): The corresponding derivatives. 
}
\label{fig3_mean_structure}
\end{figure}

Since the deformation of the sample is not accompanied by a relevant modification of the NT, one can conclude that on the atomic scale this deformation must be due to: 1) the distortion of the tetrahedral units, which can be characterized by the change in SiO bond length and the intra-tetrahedral OSiO angle; 2) rotation of the tetrahedra (assuming that the tetrahedra are rigid bodies); 3) bending motion of the inter-tetrahedral linkages (characterized by the SiOSi angle). We note that although in theory the three forms of deformation can be independent of each other, in practice they are intimately related. One expects that the coupling between (2) and (3) is stronger than the ones between (1) and (2) or (3) since changing the shape of the tetrahedral units is energetically more expensive than the rotation or bending motions and below we will show that this is indeed the case. 

In order to identify the relevant type of deformation 
we investigate in the following the related structural quantities as well as their derivatives with respect to strain.
From Fig.~\ref{fig3_mean_structure}(a)-(c) one recognizes that the change of the OSiO angle and SiO bond length is much weaker that the one of the SiOSi angle which demonstrates that the deformation of the sample is mainly due to the change of the inter-tetrahedral linkages. 
Surprisingly we find that none of the quantities presented in Fig.~\ref{fig3_mean_structure}(a)-(c) show an $\varepsilon$-dependence that mirrors the one of $E_t$, indicating that these quantities are not the relevant ones for rationalizing the strain dependence of $E_t$. However, since $E_t$ represents the slope of the SS curve, one has also to check whether $E_t(\varepsilon)$ is related to the {\it derivative} of the structural quantities.
Figures~\ref{fig3_mean_structure}(d)-(f) show that the slope $m_X=dX/d\varepsilon$ does not match the $\varepsilon$-dependence of $E_t$. (Fig.~\ref{SI_fig_struc_flex-as3} shows that this observation is independent of the nature of the modifiers.) This result can be rationalized by recalling that $E_t$ is an observable that describes the global response of the system while the considered observables $X$ are local quantities and hence the strain dependence of their average does not necessarily reflect the one of the macroscopic behavior. For instance,  the presence of a soft region in the sample may not affect the average but strongly influence the macroscopic response. Below we will discuss these mechanical heterogeneities in more detail.

\begin{figure}[htp]
\centering
\includegraphics[width=0.45\columnwidth]{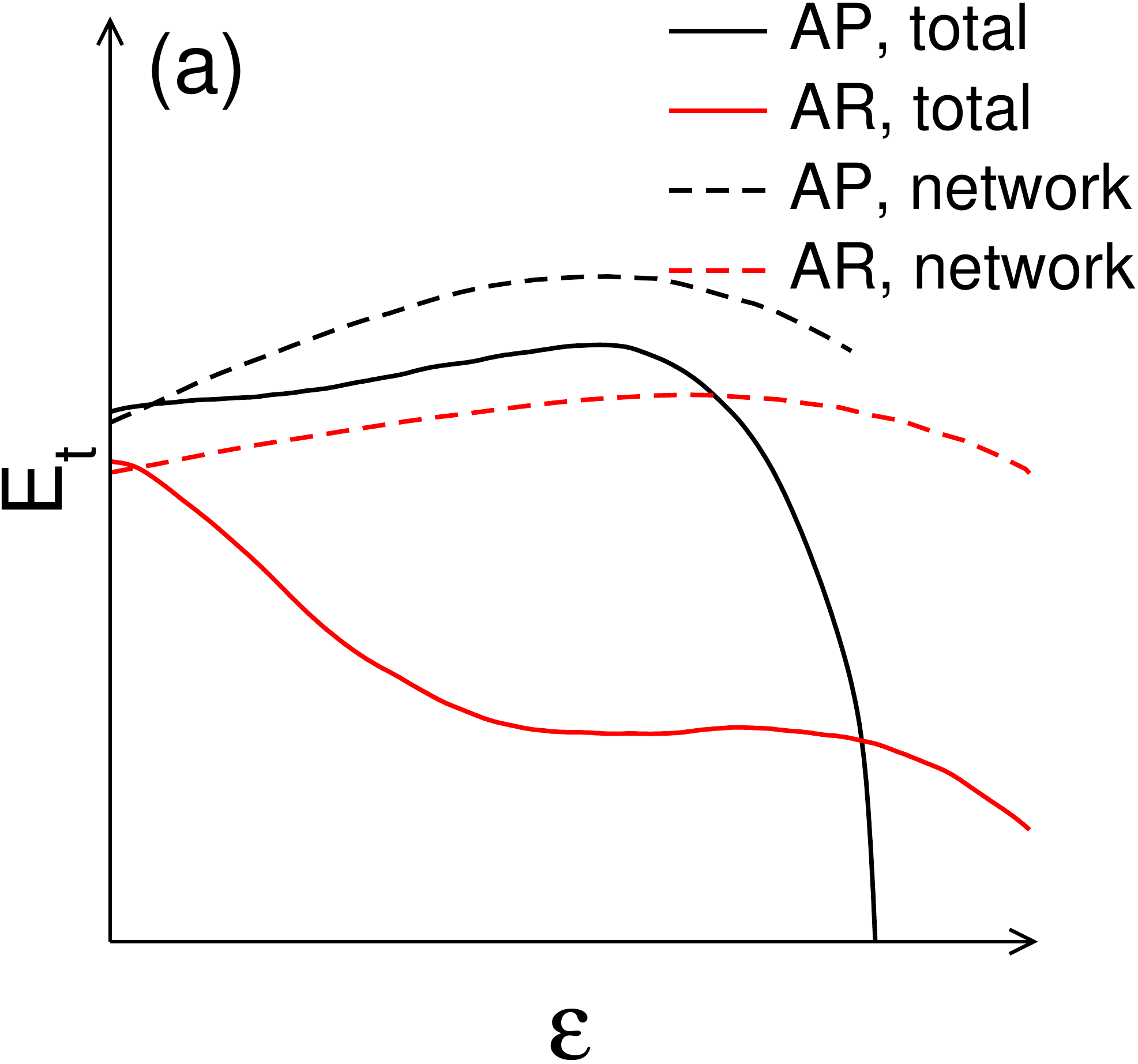}
\includegraphics[width=0.5\columnwidth]{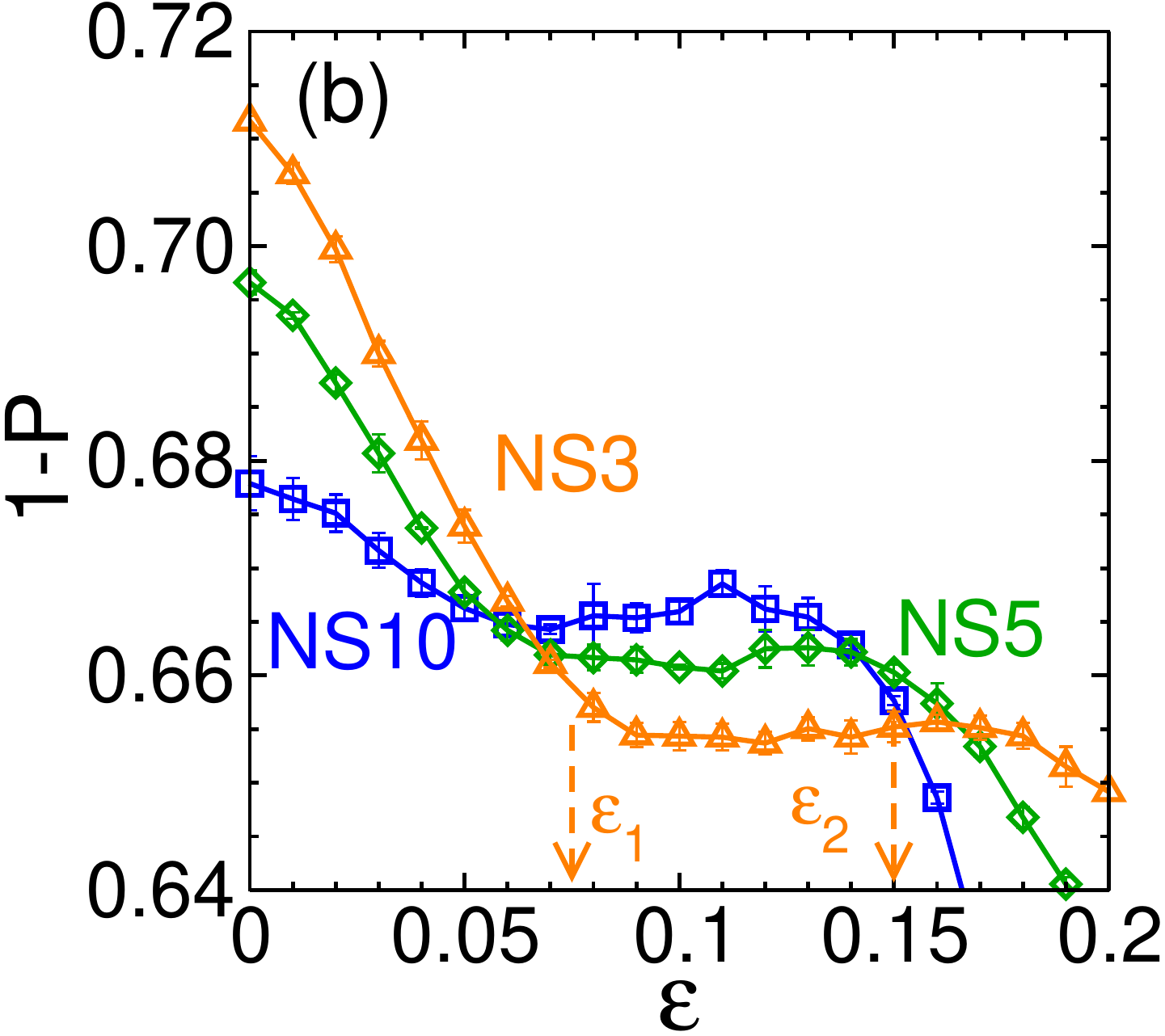}	
\includegraphics[width=0.95\columnwidth]{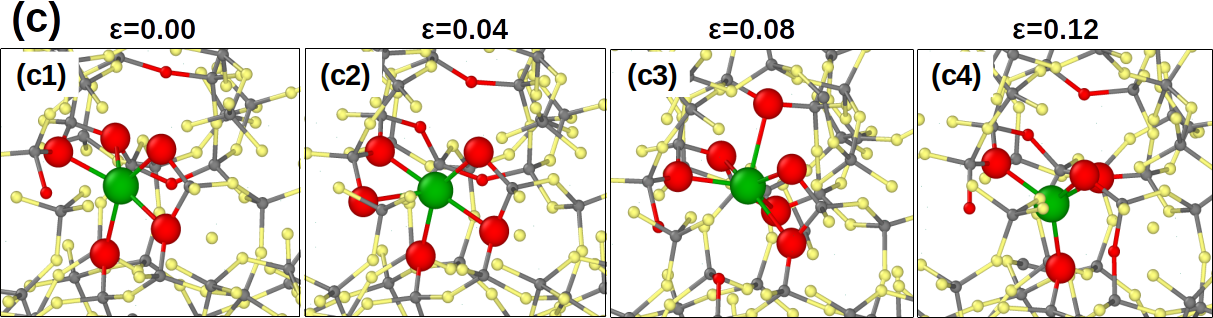}		
\caption{(a): Sketch of the strain-dependence of $E_{\rm t}$ for an alkali-poor  (AP) and an alkali-rich (AR) glass. The solid and dashed curves represent the responses of the whole sample and the Si-O network alone, respectively. (b): Probability that a Na atom will not change its bonding environment, see main text for definition. 
The two dashed lines indicate $\varepsilon_1$ and $\varepsilon_2$ for the NS3 glass. 
(c1)-(c4): Snapshots showing the change of bonding environment of a Na atom during the deformation of NS3. From left to right: $\varepsilon=$0.00, 0.04, 0.08, and 0.12. (c2)-(c4) present respectively the three possibilities for a change in the bonding of the reference Na: Unchanged $Z$ but swapped bond, increasing $Z$, and decreasing $Z$.  The atom colors depict Na (green), Si (grey), O (red and yellow). The O atoms that are currently bonded to the Na are colored red and enlarged for clarity. The bond lengths are smaller than the distances corresponding to the first minimum in $g_{\rm NaO}(r)$ and $g_{\rm SiO}(r)$.
}
\label{fig4_Na-behavior}
\end{figure}

Since neither the change of the NT nor the SiO network associated local structural quantities are able to rationalize the strain dependence of $E_t$ we turn our attention to the network modifiers. For this it is instructive
to imagine a separation of the mechanical responses into a contribution from the bare Si-O network (i.e.~assuming absence of the modifiers) and one from the modifiers.
For the case of silica, we have seen that the network rigidity, as characterized by $E_t$, increases with strain if $\varepsilon<\varepsilon_2$. It is therefore reasonable to assume that this rigidity will always increase with strain, regardless of the network connectivity. However, the slope of $E_t(\epsilon)$ of the network will depend on the modifier content since the SiO network is increasingly deteriorated by the addition of modifiers, thus making it more flexible. This idea is schematically illustrated in Fig.~\ref{fig4_Na-behavior}(a) by the two dashed lines, along with the two solid lines that represent the real response of the glasses including the effect of modifiers. A crucial step is thus to identify the microscopic mechanism which transforms the dashed lines to the solid ones.\\

Since the Si-O bonding environment is basically unchanged upon applied tension (Fig.~\ref{fig2_nc-si-bonding}), the only way the sample can acquire  additional deformability is via the motion of the Na atoms since these are weakly bonded to the SiO network and hence are mobile~\cite{horbach_dynamics_2002}.
In the following we focus on the NS$x$ glasses for elucidating the role of the modifiers.
By analyzing the coordination of the Na atoms, we find that these modifiers are indeed changing their local environment during the deformation via bond switching, a phenomenon that has been proposed to be responsible for the plasticity in other amorphous materials~\cite{zheng_electron-beam-assisted_2010,luo_size-dependent_2016,frankberg_highly_2019,lee_plasticity_2020,to2021bond}. (Fig.~\ref{SI_fig-ns3-bond-switch} shows that the coordination number ($Z$) and nearest neighbors of Na depend strongly on strain). More specifically, a Na atom can change its bonding by swapping bonds leaving $Z$ unchanged, by increasing $Z$, or by decreasing $Z$, see the snapshots in Fig.~\ref{fig4_Na-behavior}(c). 
It is thus useful to quantify the intensity of such bond switching events (BSEs) in a given sample by the probability $P$ that a Na atom changes its bonding per unit change of $\varepsilon$. 

For a better comparison with the data shown in panel (a), we show in panel (b) $1-P$, i.e.,~the probability that a Na atom is not going to change its bonding environment if strain is increased by 1\%. Interestingly, this quantity shows a non-trivial $\varepsilon$-dependence with three distinct stages that are delimited by $\epsilon_1$ and $\epsilon_2$, i.e., at the same critical strains observed for $E_t$. Also remarkable is that the height of the plateau at intermediate strain depends only weakly on the concentration of the Na atoms, which indicates that collective effects are not important for this motion.
The solid curves in panel (a) can thus be rationalized by postulating that the total response of the structure is given by the sum of the response of the Si-O network (i.e. the dashed lines in panel (a)) and of the modifiers (i.e.~the curves in panel (b)). The dependence of $E_t(\varepsilon)$ on the Na concentration can hence be attributed to the increasing influence of the Na atoms that switch bonds and thus release the local stress, leading to the presence of the plateau in $E_t$ if the modifier concentration is sufficiently large.

Figure \ref{SI_fig_bond-switch-as3} shows that
the bond switching probability depends strongly on the type of modifier in that $P$ increases with the size of the alkali atom. This increase of $P$ with alkali size can be related to the fact that atoms with larger size have a higher coordination number $Z$ which makes the change of bonding environment more probable. (See Fig.~\ref{SI_fig-as3-gr-cn} for the $\varepsilon$-dependence of $Z$ for the three types modifiers). This simple effect and the arguments presented above for the softening of the structure as a function of $P$ provide thus a microscopic explanation for the observation that $E_t$ reduces with increasing alkali size, see Fig.~\ref{fig1_AS-ss-E-bulk}(d).

\begin{figure}[htp]
\centering
\includegraphics[width=0.95\columnwidth]{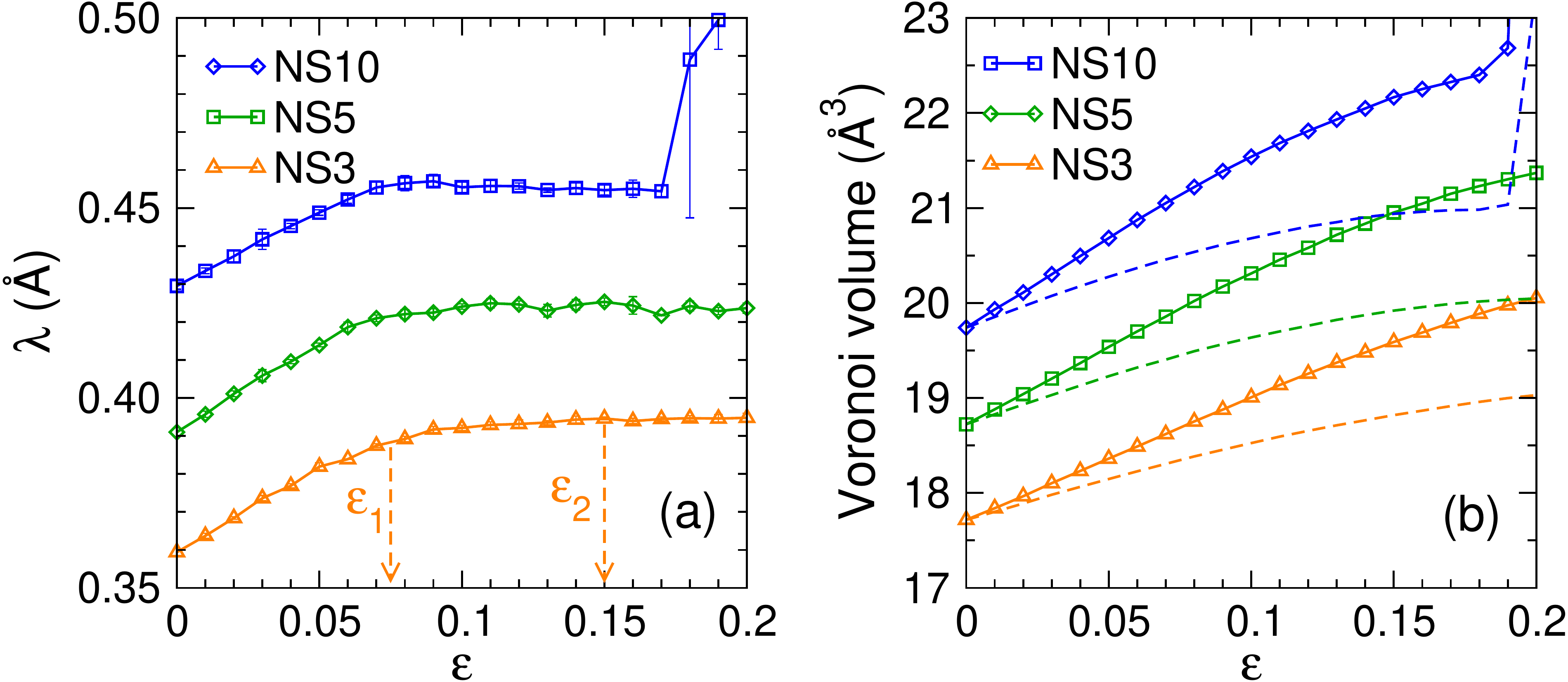}
\caption{(a) Mean vibration displacement of the Na atoms, see the main text for definition. The dashed lines indicate $\epsilon_1$ and $\epsilon_2$ for NS3. (b) Atomic volume of Na estimated by Voronoi tessellation. The dashed lines are for affine deformation.
}
\label{fig5_mechanism-Na-plateau}
\end{figure}

In Fig.~\ref{fig4_Na-behavior}(b) the rapid decrease of $1-P$, i.e., the intensity of the BSEs increase, at large strains ($\varepsilon>\varepsilon_2$) reflects simply the yielding of the glass structure on the global scale. In contrast to this, the plateau in the $\varepsilon$-dependence of $P$ is surprising and its origin needs to be clarified. 
Since we have seen that the BSEs are due to the motion of the alkali atoms and not of the network, it is useful to probe their motion in more detail.
Figure~\ref{SI_fig_nsx-distri-nonaffinedisp-A} demonstrates that during the tension most of the Na atoms move only by a small distance, i.e.~non-affine displacement, and that the probability for a Na atom to escape from its cage occupied at $\epsilon=0$ is at most a few percent up to the critical strain $\epsilon_2$. Thus it is reasonable to view the BSEs as discussed above to be basically of vibrational nature.  We have therefore determined the mean vibrational displacement $\lambda$ of the modifier atoms which is proportional to the size of the cage within which the atoms vibrate locally (see Methods).
Figure~\ref{fig5_mechanism-Na-plateau}(a) shows that $\lambda$ first increases before saturating at a value of the strain that is close to $\varepsilon_1$. This saturation is not trivial since the locally available space for the Na atoms, characterized by their Voronoi volume, increases monotonously with strain, see panel (b). (This increase is directly related to the fact that for these glasses the Poisson's ratio is smaller than 0.5, see Fig.~\ref{SI_fig_nsx-as3-poisson-ratio}.)
Hence we conclude that for $\varepsilon < \varepsilon_1$ the increasing atomic volume allows the Na atoms to vibrate with a larger amplitude but that above a certain volume this amplitude has reached its maximum value. This increasing amplitude allows the Na atom to change bonding partners and therefore $\lambda$ shows a similar $\varepsilon$-dependence as $P$, see Fig.~\ref{fig4_Na-behavior}(b), which in turn leads to the decrease of $E_t$. Once the maximum amplitude is reached, the intensity of the BSEs becomes independent of $\varepsilon$, and together with the weak $\varepsilon$-dependence of the network rigidity (Fig.~\ref{fig4_Na-behavior}(a)), $E_t$ becomes nearly a constant. 

In the context of Fig.~\ref{fig3_mean_structure} we have argued that the plateau in $E_t$ is not seen in the mean value of the structural quantities because the mechanical response will be strongly influenced by the structural heterogeneities of the sample. To probe the effect of these heterogeneities we  determine how a local structural quantity $X$ (angle, bond length, etc.) changes per unit change of $\varepsilon$, giving $\Delta X$. We find that for all cases considered, the distribution of $\Delta X$ is described well by a Gaussian, see Fig.~\ref{SI_fig_nsx-struc-distri}, and therefore its standard deviation $\chi_X$ is a good measure of the heterogeneity of the response of the system.

\begin{figure}[ht]
\includegraphics[width=0.95\columnwidth]{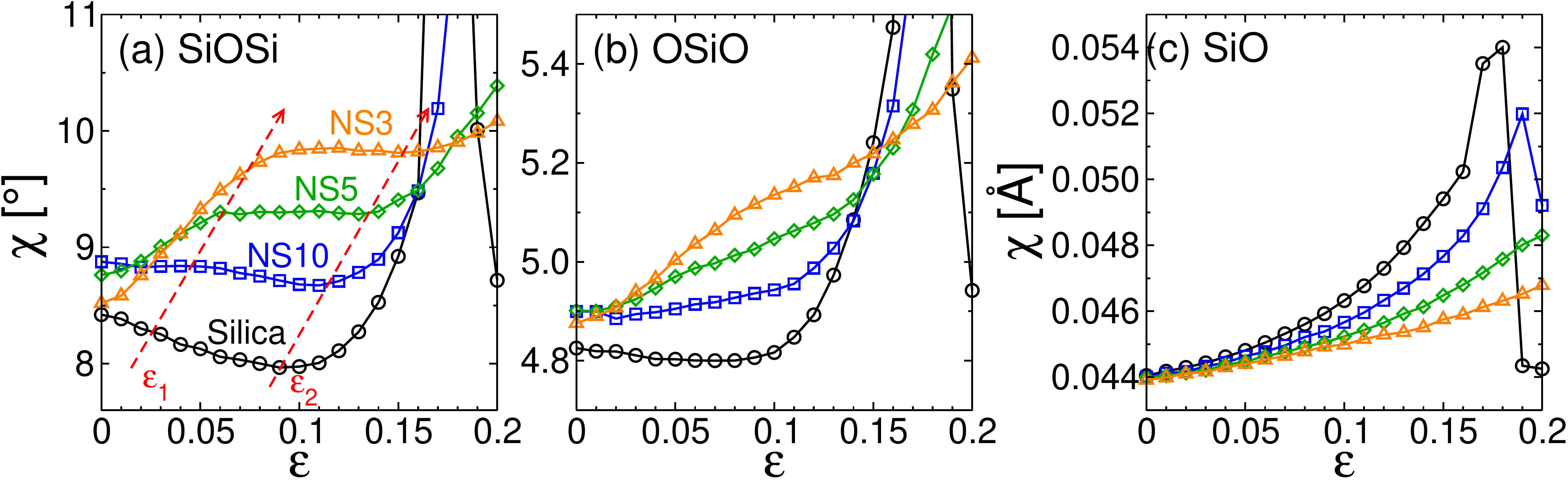}
\caption{ Standard deviation of the incremental structural quantities. (a): SiOSi angle; (b) OSiO angle; (c) SiO bond length. In panel (a), $\varepsilon_1$ and $\varepsilon_2$ are indicated by arrows. 
}
\label{fig6_struc_flex}
\end{figure}

In Fig.~\ref{fig6_struc_flex} we present $\chi_X$ for different quantities $X$. Remarkably, one sees that the $\varepsilon$-dependence of $\chi_{\rm SiOSi}$, panel (a), matches very well the one of the $E_t$, i.e.~a nearly perfect anti-correlation between the two quantities is observed, irrespective of the composition. (Figure~\ref{SI_fig_struc_flex-as3} shows that this good correspondence is unaffected by changing the alkali species.) For silica, the network is fully connected and becomes increasingly rigid with strain ($E_t$ increases) and this is accompanied by a decrease of $\chi_{\rm SiOSi}$, i.e. the system becomes more homogeneous. However, the stiffening of the network reaches a limit at around 10\% strain, beyond which the structure will yield globally. Consequently, with further increasing strain the structure softens, $E_t$ decreases, and $\chi_{\rm SiOSi}$ increases quickly. With the addition of Na the network becomes less connected and hence easier to deform. Upon the application of stress the soft regions will expand more than the rigid ones, triggering the BSE and hence relax the local stresses, i.e., the system becomes more heterogeneous and thus the $\chi$ increases. Once the soft regions have undergone this expansion (in accordance with the increase of atomic volume and vibrational amplitude as seen in Fig.~\ref{fig5_mechanism-Na-plateau}) the bonding structure becomes independent of $\varepsilon$ and hence $\chi$ becomes constant. One thus finds in the strain dependence of $\chi$ the various stages marked by $\epsilon_1$ and $\epsilon_2$ in Fig.~\ref{fig1_AS-ss-E-bulk}(b). 

It is also interesting to examine the values of $\chi_{\rm SiOSi}$ at $\varepsilon=0$. Starting from silica one finds that a small addition of modifiers makes the structural response more heterogeneous (the case of NS10). However, further addition of modifiers tend to homogenize again this response (the case of NS5 and NS3), i.e.,~$\chi$ decreases again. Thus $\chi$ can be used directly to probe the importance of the heterogeneities. This data also demonstrates that a small $\chi$ is not necessarily associated with a large $E_t$. This can be rationalized by the fact that $E_t$ is related not only to the heterogeneous response of the local structure but also to the overall three-dimensional connectivity of the network which depends strongly on the composition.  

From Fig.~\ref{fig6_struc_flex}(b) one sees that  $\chi_{\rm OSiO}$ shows the same qualitatively  $\varepsilon$-dependence as $\chi_{\rm SiOSi}$. However, the relative change of $\chi_{\rm OSiO}$ is significantly smaller than the one found in $\chi_{\rm SiOSi}$, indicating that the $\varepsilon$-dependence of the heterogeneity in the OSiO angular response is mainly due to the coupling of this angle with the SiOSi angle. The $\varepsilon$-dependence of $\chi_{\rm SiO}$, panel (c), is qualitatively very different from the ones for the angles in that the heterogeneity increases monotonously for all compositions, indicating  that this quantity is not relevant for the dependence of the flexibility of the network on composition. These results support our argument made above that the distortion of the tetrahedra is energetically costly compared to other deformation modes, in agreement with the fact that in the vibrational density of states intra-tetrahedral vibrational modes (mainly stretching) are at higher frequencies than the inter-tetrahedral ones (bending and rocking)~\cite{taraskin_nature_1997}.
We thus have strong evidence that the heterogeneous response of the inter-tetrahedral linkages, influenced by the unusual dynamical behavior of the modifiers, is the primary structural manifestation of the non-linear elastic properties of the glasses.

In conclusion, we have demonstrated that for the case of silicate glasses the presence of heterogeneities in the local structural properties gives rise to an unexpected behavior of the non-linear elastic behavior of such systems. This behavior is intimately related to the presence of species (here the alkali atoms) that are more mobile than the atoms forming the matrix (Si-O network) since they allow to react locally to the applied stress. 
Although we are not aware of any experimental results regarding this unexpected mechanical behavior, its detection should be possible by carefully probing the elastic properties of alkali-silicate glasses at large tensile strains.
Note that although the details regarding the decoupling of the response between mobile and immobile parts of the sample will depend on the system considered, it can be expected that the identified mechanism is in fact very general for materials which consist of atomic species with distinctive mobilities.  Our findings can thus be expected to trigger further studies on the deformation behavior of complex materials and should also be of practical relevance for the design of materials with tailored mechanical properties.

\section{ACKNOWLEDGMENTS}
We thank P.~K. Gupta for discussions. ZZ acknowledges a grant from China Scholarship Council (NO. 201606050112). WK is member of the Institut Universitaire de France.
The simulations were done by utilizing the HPC resources
of CINES under the allocation A0050907572 and A0070907572
attributed by GENCI (Grand Equipement National de Calcul
Intensif) and the computational resources at XJTU.\\

Author contributions: ZZ, SI, and WK designed the research. ZZ carried out the simulations. ZZ and WK interpreted the data and wrote the paper.

Competing interests: The authors declare no competing financial
interests. 

Data and materials availability: All data in the manuscript
or the Materials are available from W. Kob upon reasonable request.

\normalem  

\bigskip
\noindent{\Large \bf Methods}\\
\noindent{\bf Systems and simulations}. 
We performed MD simulations for pure SiO$_2$, binary Na$_2$O-$x$SiO$_2$ with $x=3, 5, 10$, and binary A$_2$O-3SiO$_2$ with A=Li, Na, K. The atomic interactions are described by a two-body effective potential~(SHIK)~\cite{sundararaman_new_2018,sundararaman_new_2019}
which has been shown to give a reliable description of the
structural and mechanical properties of sodium silicate glasses~\cite{zhang_potential_2020,zhang2020jcp}. Our samples contain typically $6\times10^5$ atoms, corresponding to a cubic box sizes of side length around 20~nm, which is sufficiently large to avoid noticeable finite size effects~\cite{zhang_potential_2020}. Periodic boundary conditions were applied in all directions. The samples were first melted and equilibrated at 3000~K and then cooled  down to 300~K at zero pressure ($NPT$ ensemble) with a cooling rate of 0.25~K/ps. The glass samples were annealed at 300~K for 160~ps and then subjected to uniaxial tension with a constant strain rate of 0.5~ns$^{-1}$. We emphasis that the simulation parameters were chosen the results presented in this paper do not depend significantly on these parameters (see discussion in Refs.~\cite{zhang_potential_2020,zhang_thesis_2020}). 

Temperature and pressure were controlled using a Nos\'e-Hoover thermostat and barostat~\cite{nose_unified_1984,hoover_canonical_1985,hoover_constant-pressure_1986}. All simulations were carried out using the Large-scale Atomic/Molecular Massively Parallel Simulator software (LAMMPS)~\cite{plimpton_fast_1995} with a time step of 1.6~fs. 
The results presented in this manuscript correspond to one melt-quench sample for each composition. However, we emphasize that the system sizes considered in this study are sufficiently large to make sample-to-sample fluctuations negligible as long as the samples do not start to fracture. The error bars were estimated as the standard deviation from 3 simulations in which the samples were put under tension in the 3 different axial directions.\\

\noindent{\bf Vibrational amplitude}. We characterize the motion of the atoms by using the vibrational displacement $\lambda$ defined as  
$\lambda_i=\sqrt{\langle [\vec{r}_i(t)-\vec{r}_i(0)]^2 \rangle_{\tau_0}}$
, where $\vec{r}_i(t)$ is the position of atom $i$ at time $t$ and $\tau_0$ is the time interval over which we take the average. We have chosen $\tau_0$ to be 0.4-0.5~ps since this is the time scale that corresponds to the appearance of the plateau in the mean-squared displacement (see Fig.~\ref{SI_fig_msd} in the SI). To measure the MSD or vibrational displacement we take a configuration at a given strain and then carry out a simulation at 300~K in the $NVT$ ensemble for several tens of picoseconds during which the MSD is determined. 

\onecolumngrid

\clearpage
\noindent{\Large \bf
Supplemental Information for: \\ 
Origin of the non-linear elastic behavior of silicate glasses}\\

\noindent Zhen Zhang$^{1,2}$, Simona Ispas$^2$, and Walter Kob$^{2*}$\\
\noindent $^1${\it Center for Alloy Innovation and Design, State Key Laboratory for Mechanical Behavior of Materials, Xi’an Jiaotong University, Xi’an 710049, China}\\
\noindent $^2${\it Laboratoire Charles Coulomb (L2C), University of Montpellier and CNRS, F-34095 Montpellier, France}\\[5mm]
Corresponding author: walter.kob@umontpellier.fr

\renewcommand{\figurename}{Figure}
\renewcommand{\thefigure}{S\arabic{figure}}
\setcounter{figure}{0}

\bigskip

\noindent{\bf 1. On the reliability of the interaction potential}

\begin{figure}[ht]
\includegraphics[width=0.95\columnwidth]{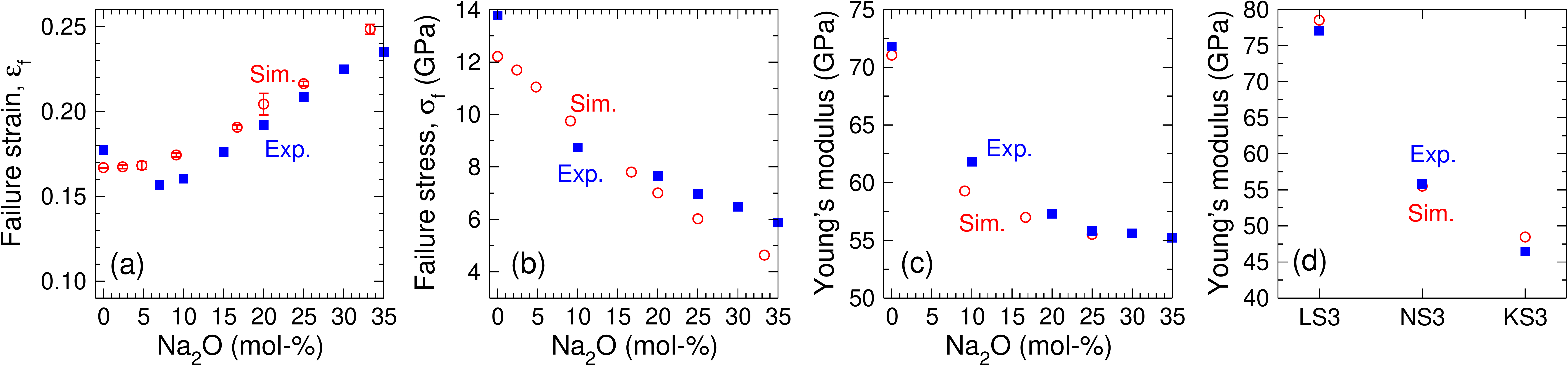}
\caption{(a) Failure strain of NSx glasses. (b) Failure stress of NSx glasses. (c) Young's modulus of NSx glasses. (d) Young's modulus for AS3 glasses. In (a)-(c) the experimental data are taken from Ref.~\cite{lower_inert_2004} and were measured by using a two-point bending technique on a series of flaw-free glass fibers. Note that the failure stress was calculated by using an expression which describes well the non-linear elastic behavior of silicate glasses~\cite{gupta_intrinsic_2005}. In (d) the experimental data are taken from Ref.~\cite{bansal_handbook_1986}. In (b), (c), and (d) the error bars for the simulation data are smaller than the symbol size.  \\
}
\label{SI_fig_failure-Y-compare}
\end{figure}

In Fig.~\ref{SI_fig_failure-Y-compare} we compare the
various mechanical quantities as predicted by our simulations with the corresponding experimental data.
The latter were measured by using a two-point bending technique on a series of flaw-free glass fibers~\cite{lower_inert_2004,bansal_handbook_1986}, thus representing the intrinsic properties of the glasses. (Note that the experimental value of the failure stress, panel (b), was calculated by using an empirical expression which describes well the non-linear elastic behavior of silicate glasses~\cite{gupta_intrinsic_2005}.) It can be seen that our simulated results agree well with the experimental results not only in the elastic modulus  (at $\epsilon=0$) but also the failure strength and strain, indicting that the potential we have used for our simulations is reliable. 

In Fig.~\ref{fig1_AS-ss-E-bulk} of the main text we have documented that the strain dependence of $E_t$ shows for alkali-rich glasses, e.g.~NS3, an unexpected plateau at intermediate values of $\varepsilon$. In order to test whether this result is not a strange peculiarity of the interaction potential used for the simulations, we have carried out simulations for the NS3 system using other potentials: In addition to the SHIK potential, we also consider the potential by Habasaki and Okada (HO)~\cite{habasaki_molecular_1992} and a reactive force field  (reaxFF)~\cite{hahn_development_2018}, both of which are popular choices for simulation studies of the mechanical properties of sodium silicate glasses. Our simulation tests were divided into two stages: In the first stage, we used potential $A$ to produce the glass via a melt-quench procedure. In the second stage, we switched to potential $B$, annealed first the prepared glass sample for 160~ps and subsequently the sample was put under tension until it fractured. The samples contained around 40000 atoms which correspond to a cubic box of side length $L=8$~nm. All simulations were performed using the $NPT$ ensemble with zero pressure, and the fracture simulations were done at 300~K with a constant strain rate of 0.5/ns.    

The reason that we did not use the reaxFF potential for the melt-quench procedure is that  simulations with this potential are computationally very expensive. Instead, we choose to use this potential to relax (at 300~K) the glass structure as generated by using other potentials. This strategy has been successfully adopted in previous simulation studies which demonstrated that the relaxed glass structure will be noticeably different from the initial structure~\cite{yu_reactive_2017}.

\begin{figure}[htp]
\centering
\includegraphics[width=0.9\columnwidth]{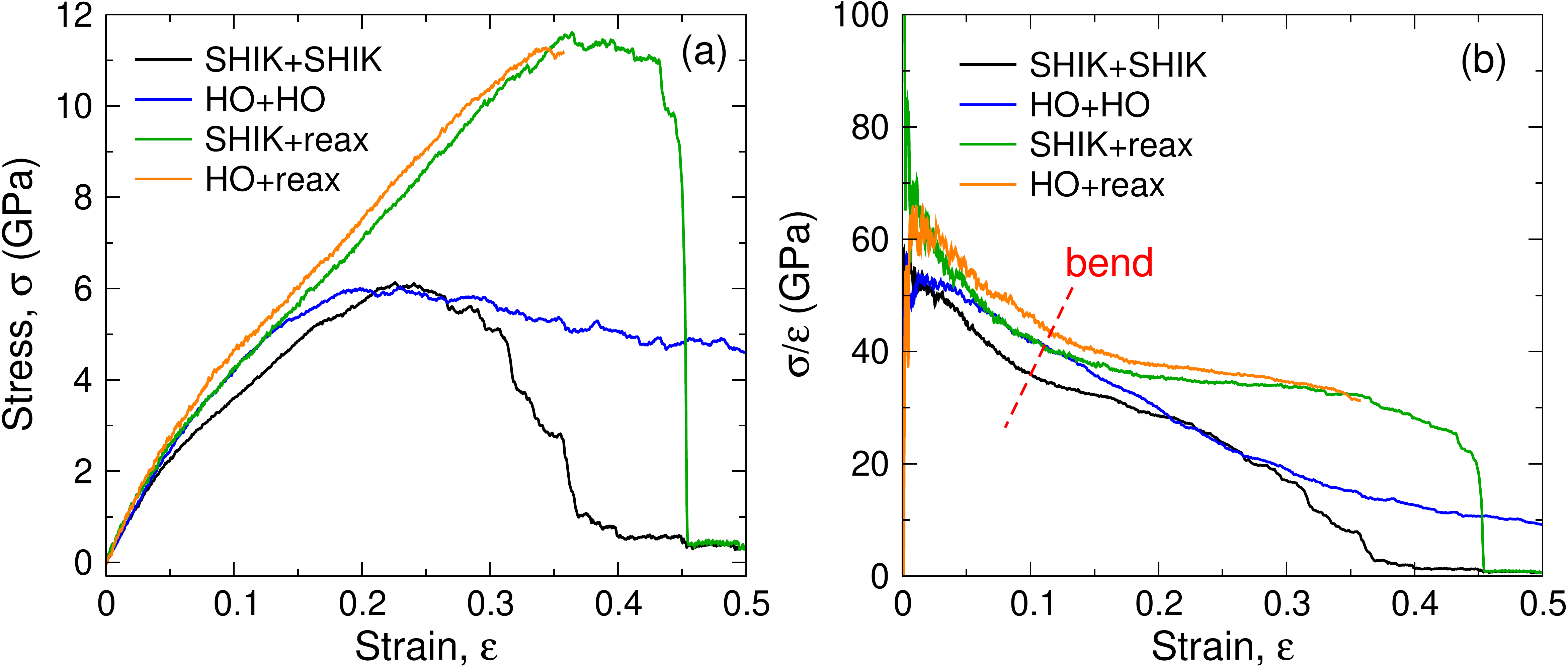}
\caption{Influence of interaction potential on the deformation of the NS3 glass. (a) Stress-strain curve. (b) The ratio between stress and strain, $\sigma/\varepsilon$. Notation of the legend: $A+B$: $A$ is the potential used for producing the glass via a melt-quench process and $B$ is the potential used for the fracture simulation.}
\label{SI_fig_ns3-bend-mix-pot}
\end{figure}

\begin{figure}[ht]
\includegraphics[width=0.55\columnwidth]{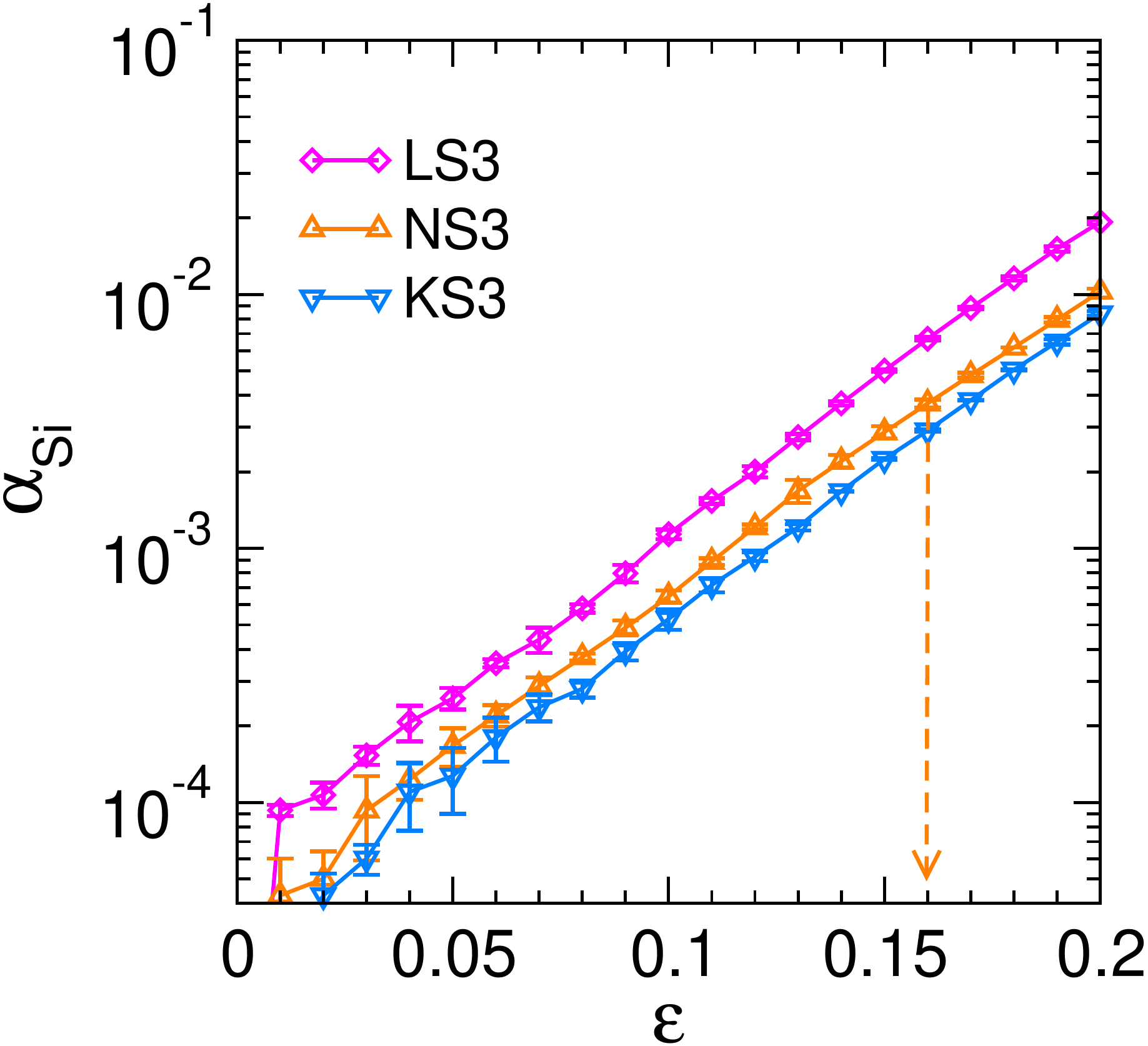}
\caption{Fraction of Si atoms that have changed their bonding environment, $\alpha_{\rm Si}$=$N_{\rm Si'}$/$N_{\rm Si}$, where $N_{\rm Si'}$ is the number of Si atoms which have at least one of their neighbors changed with respect to the initial configuration ($\epsilon=0$). The vertical arrow indicates approximately  $\varepsilon_2$ of NS3. 
}
\label{SI_fig_nc-si-bonding-as3}
\end{figure}

\begin{figure}[htp]
\centering
\includegraphics[width=0.92\columnwidth]{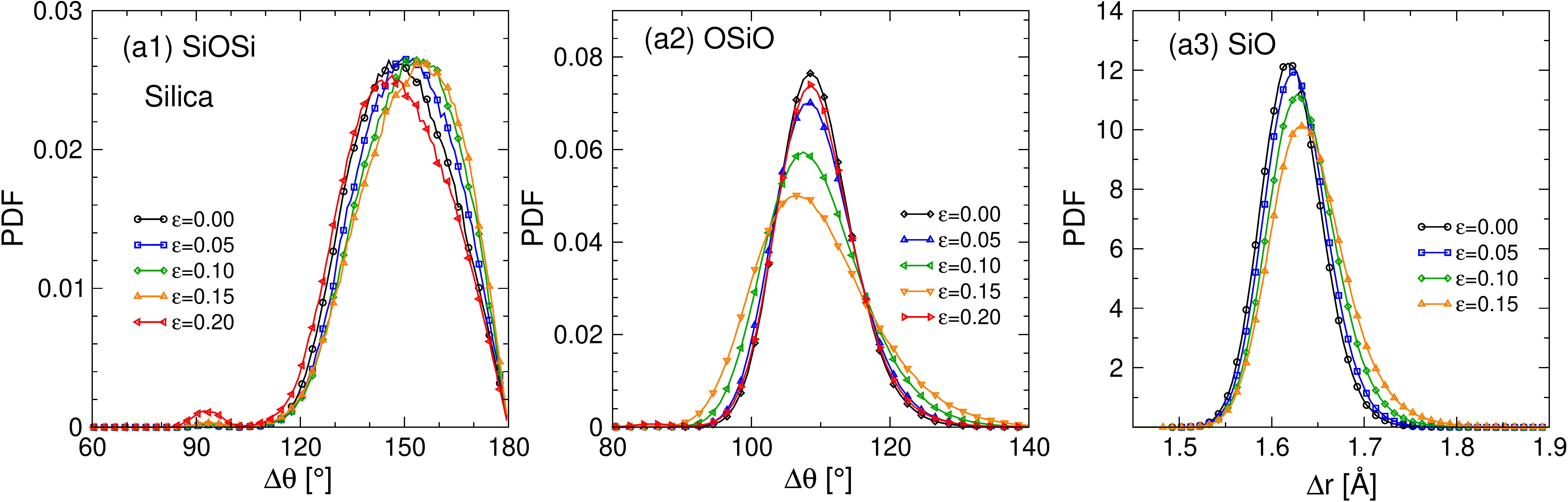}
\includegraphics[width=0.92\columnwidth]{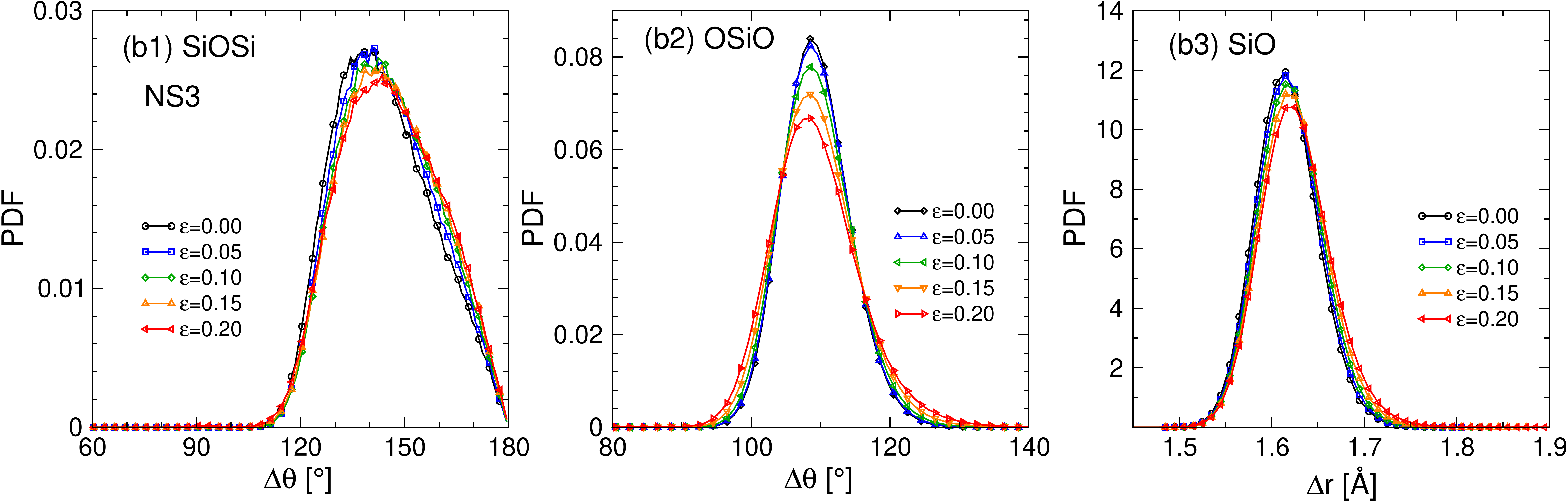}
\includegraphics[width=0.92\columnwidth]{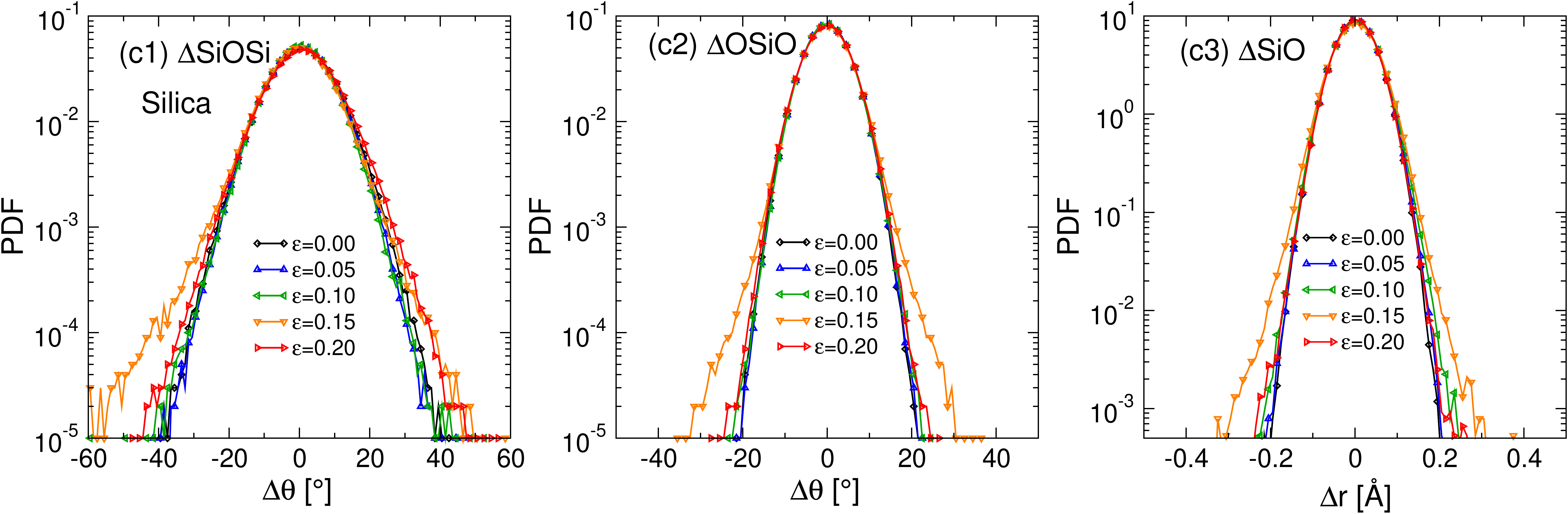}
\includegraphics[width=0.92\columnwidth]{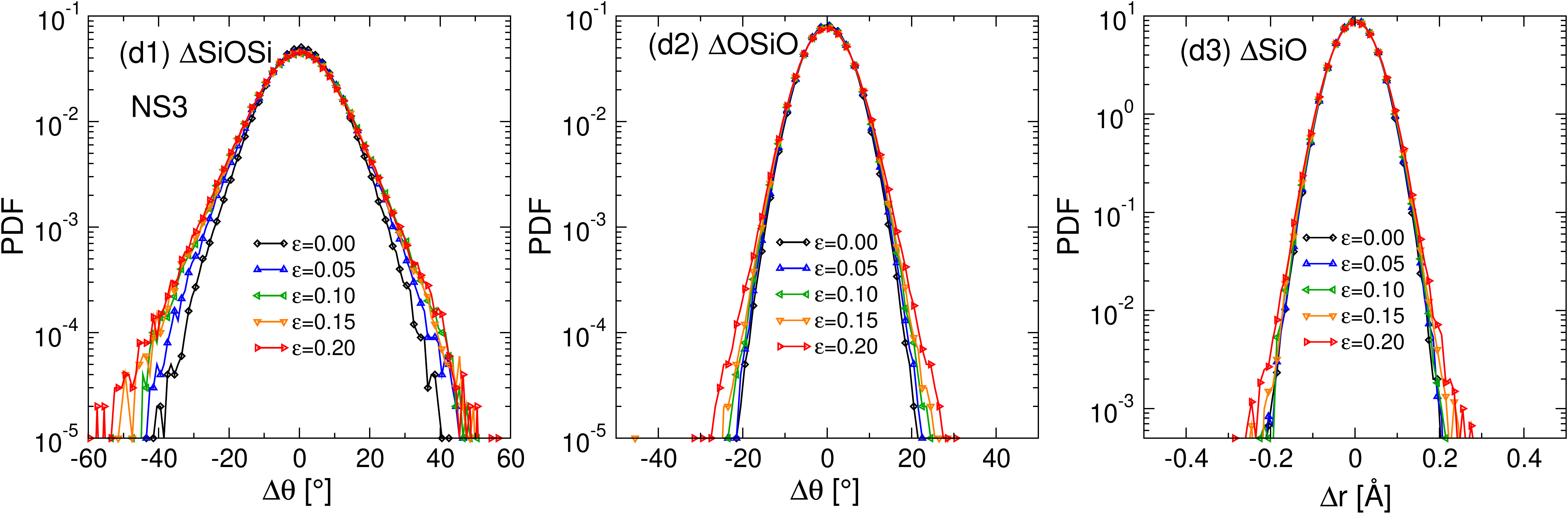}
\caption{Distribution of various structural quantities. (a1)-(a3) are for the angles and bond length of silica. (b1)-(b3) are for the angles and bond length of NS3. Silica shows a stronger dependence on the strain than NS3, which can be attributed to its more constrained network. Rows (c1)-(c3) and (d1)-(d3) are for the incremental change of the structural quantities with a strain variation of 2\%. In (c1-d3),  all quantities exhibit Gaussian-like distribution with a mean value close to zero. 
}
\label{SI_fig_nsx-struc-distri}
\end{figure}

\begin{figure}[ht]
\includegraphics[width=0.95\columnwidth]{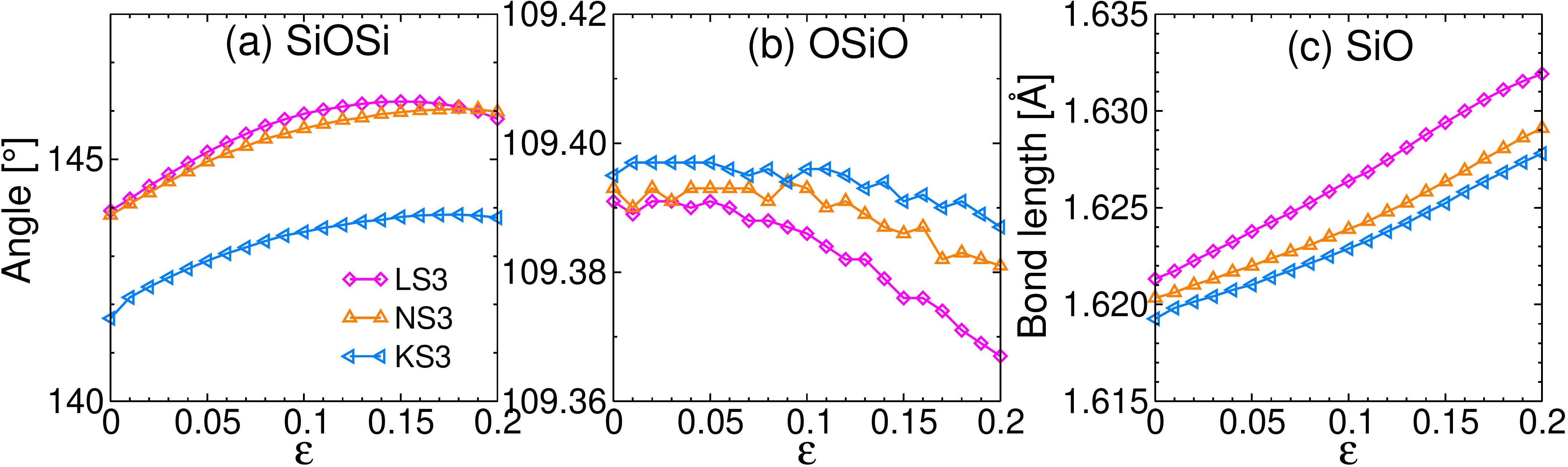}
\includegraphics[width=0.95\columnwidth]{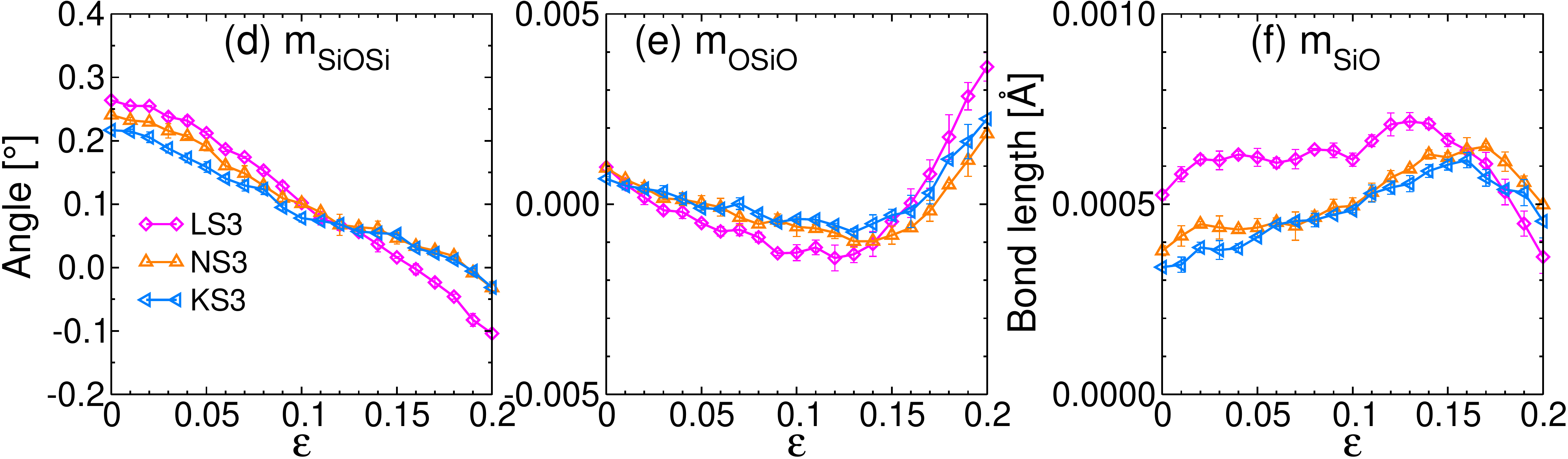}
\includegraphics[width=0.95\columnwidth]{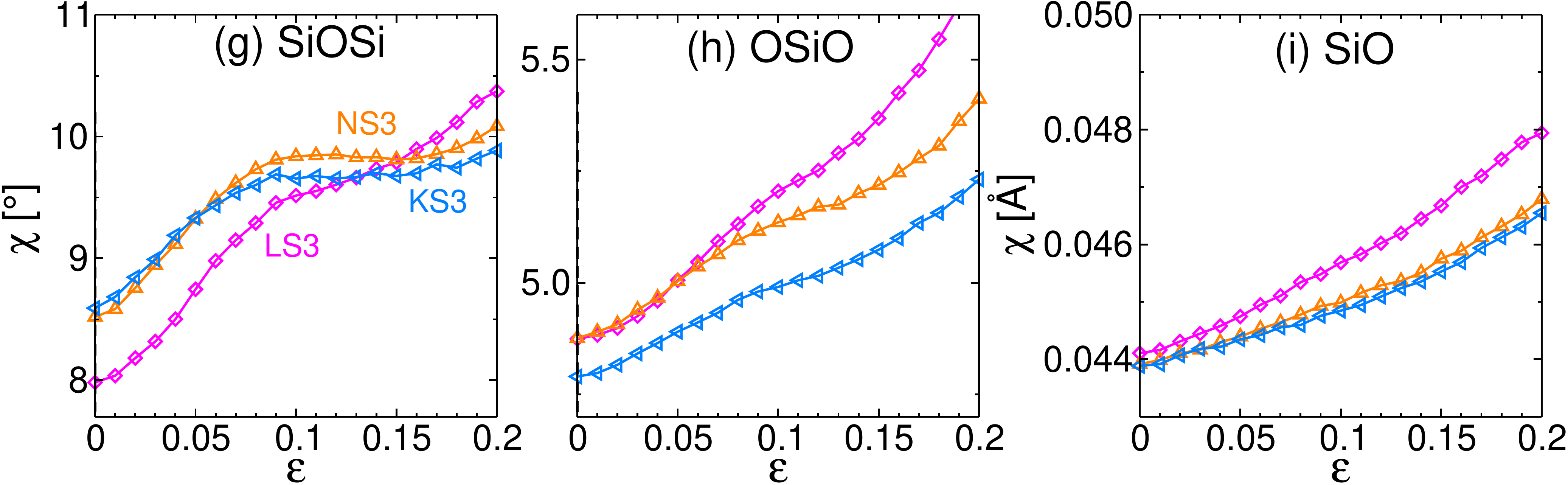}
\caption{Structural quantities that characterize various deformation modes of the Si-O network for the AS3 glasses. (a)-(c): The mean values. (d)-(f): Derivative of the data in panels (a)-(c). 
(g)-(i): Standard deviation of the incremental structural quantities. 
}
\label{SI_fig_struc_flex-as3}
\end{figure}

\begin{figure}[ht]
\includegraphics[width=1\columnwidth]{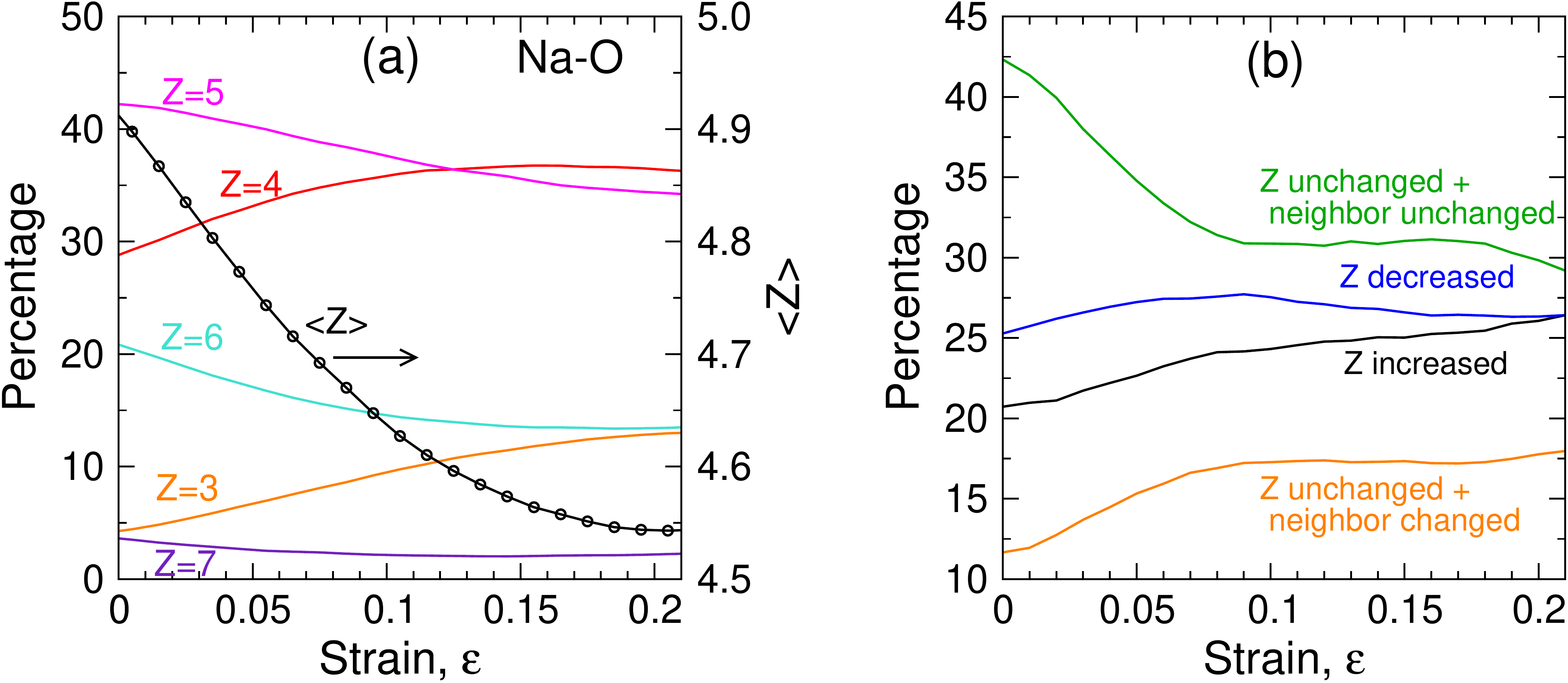}
\caption{(a) Distribution of the coordination number $Z$ of Na in the NS3 glass. The cutoff distance is chosen to be the first minimum of $g_{\rm NaO}(r)$, i.e.~2.95 \AA. The mean coordination number is shown on the right ordinate (black solid line with symbols).  (b) Incremental ($\Delta \varepsilon=0.01$) changes in Na-O bonding during tensile loading. 
For the Si-O bonding, the percentage of $Z=4$ is more than 99.9\% and no appreciable amount of bond switching events can be detected. 
}
\label{SI_fig-ns3-bond-switch}
\end{figure}

\begin{figure}[htp]
\centering
\includegraphics[width=0.5\columnwidth]{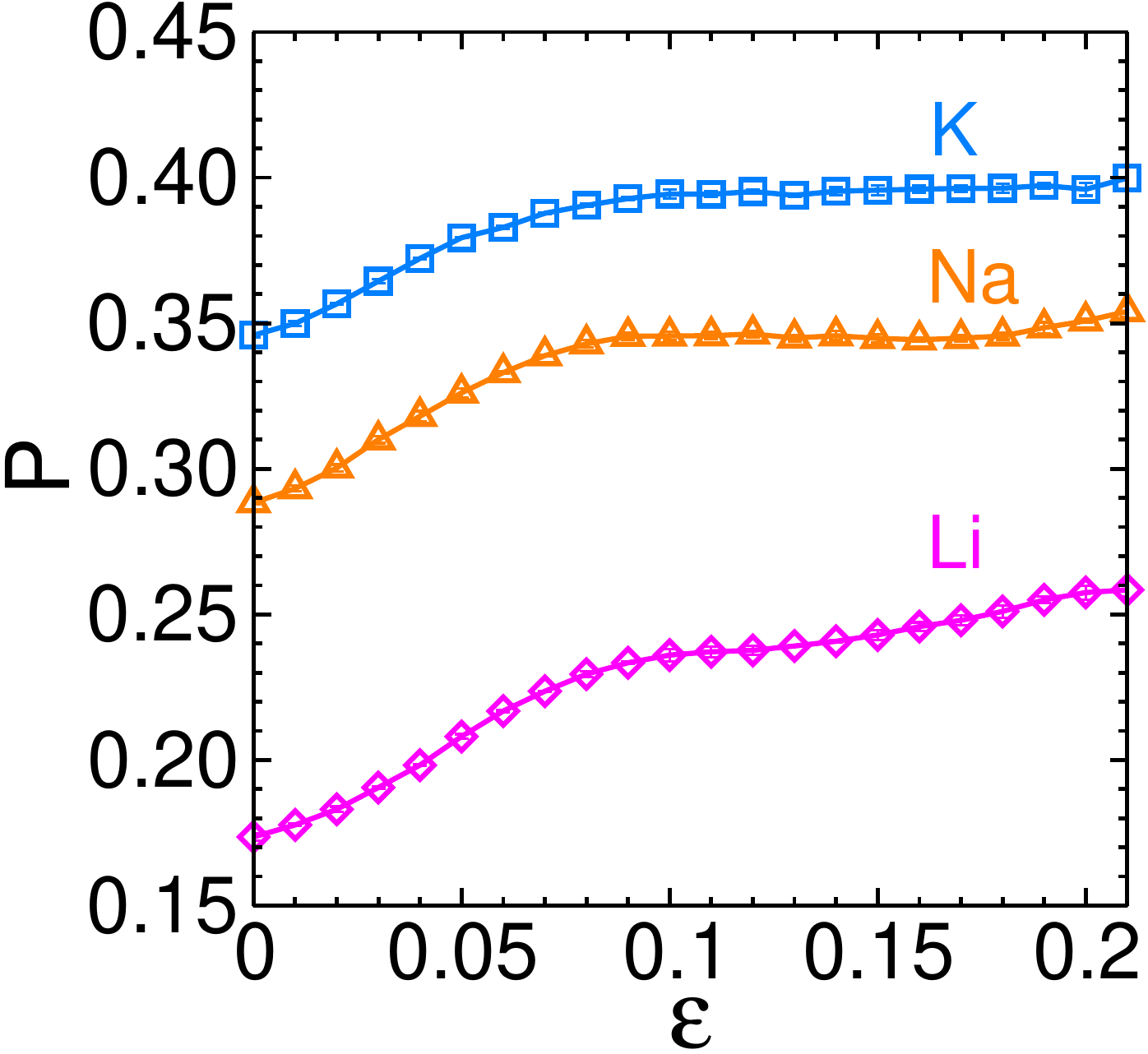}	
\caption{Probability that a modifier in an AS3 glass will change its bonding environment, see the main text for definition. Large alkali atoms are more likely to change their bonding partners. }
\label{SI_fig_bond-switch-as3}
\end{figure}

\begin{figure}[ht]
\includegraphics[width=0.8\columnwidth]{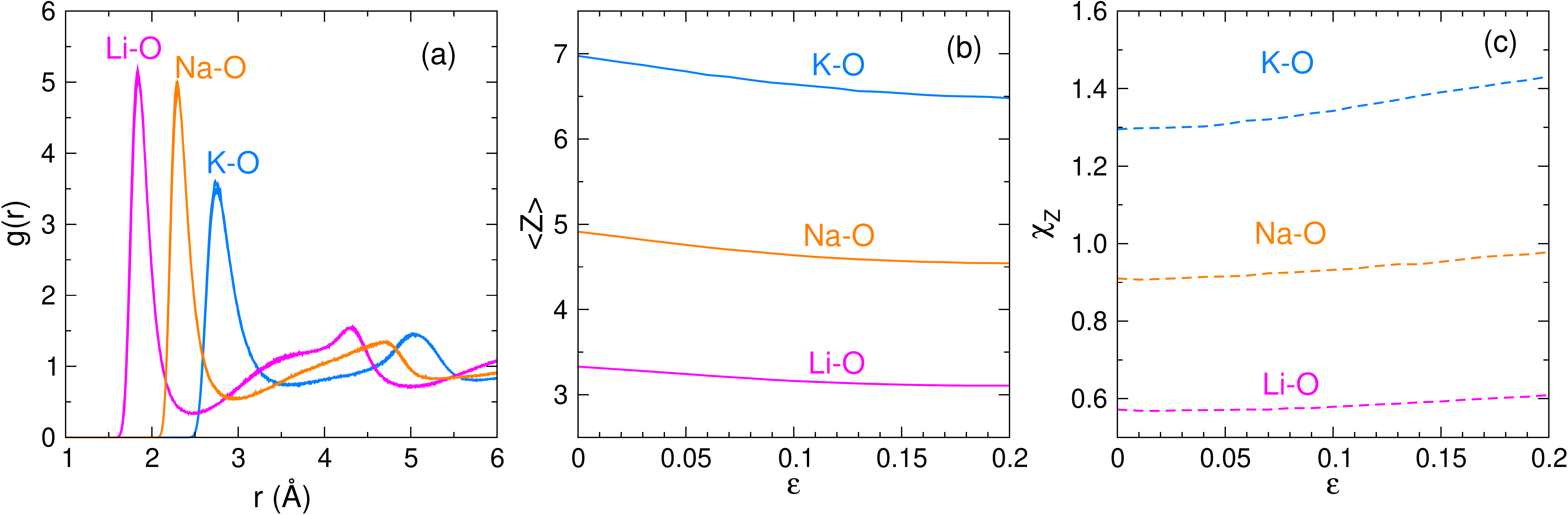}
\caption{(a) Radial distribution function for the A-O pairs of the AS3 glasses at zero strain. (b) Mean coordination number of the alkali atoms. (c) The standard deviation of the coordination number $\chi_Z$. The cutoff distances are chosen to be the location of the first minimum of $g(r)$ shown in panel (a), which depends very weakly on strain.  
With increasing size of the modifier, the coordination number becomes larger, panel (b), and also the distribution of $Z$ becomes broader, panel (c).}
\label{SI_fig-as3-gr-cn}
\end{figure}

\begin{figure}[htp]
\centering
\includegraphics[width=0.95\columnwidth]{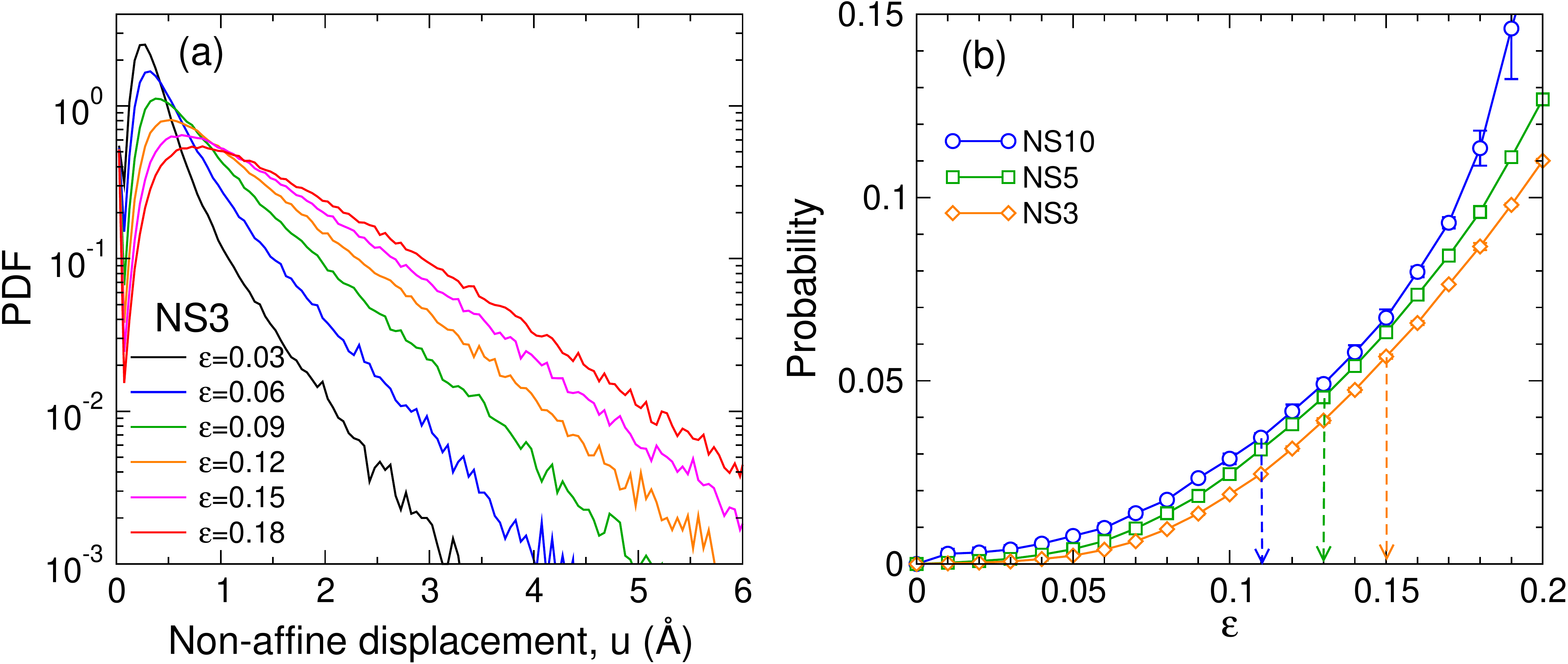}
\caption{(a) Distribution of the non-affine part of the displacement of Na, $u$, with respect to the initial configuration. Note that $u=\sqrt{D_{\rm min}^2}$, where $D_{\rm min}^2$ is the squared non-affine displacement introduced in Ref.~\cite{falk_dynamics_1998} which evaluates the nonlinear movement of atom $i$ with respect to its $N_i$ neighboring atoms. The value of $D_{\rm min}^2$ is normalized by $N_i$. The cutoff for evaluating $D_{\rm min}^2$ is chosen to be 3.0~\AA, i.e.~the distance corresponding to the first minimum of $g_{\rm NaO}(r)$. (b) The probability of Na atoms to have a non-affine displacement $u>3.0$~\AA, i.e.~atoms are likely to have escaped from their cages defined at $\varepsilon=0$. The arrows indicate $\varepsilon_2$ and one sees that the probability of Na escaped from their cages are small ($P<6$\% at most) up to this critical strain.
}
\label{SI_fig_nsx-distri-nonaffinedisp-A}
\end{figure}

\begin{figure}[htp]
\centering
\includegraphics[width=0.75\columnwidth]{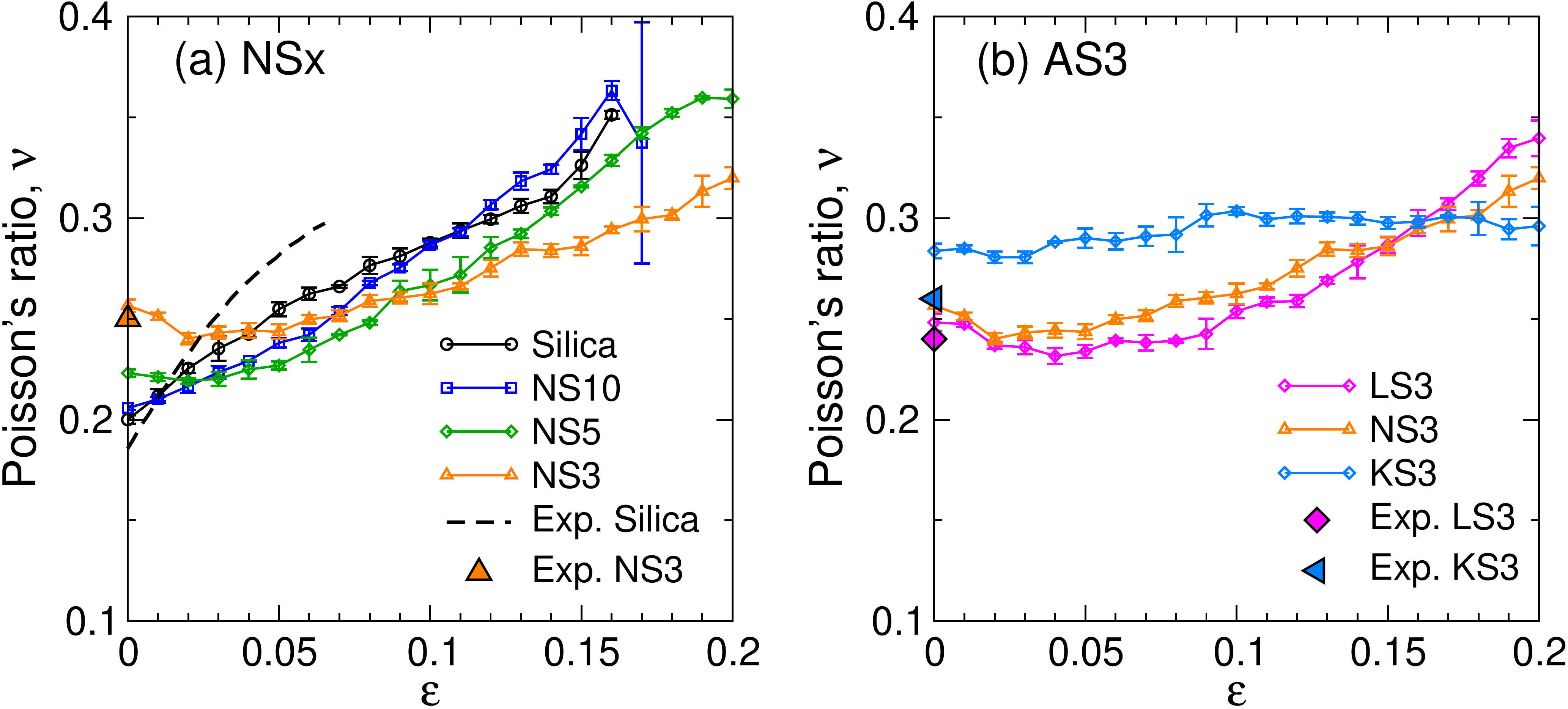}
\caption{ (a-b) Poisson's ratio $\nu$ for the NSx and AS3 glasses, respectively.  An increasing trend is observed for all compositions. The $\varepsilon$-dependence seems to become weaker upon increasing the concentration and the size of the modifiers. The symbols at $\varepsilon=0$ are experimental data~\cite{bansal_handbook_1986,januchta_elasticity_2019}. The dashed line in panel (a) is a fit to the experimental data of silica glass fibers up to 7\% strain~\cite{guerette_nonlinear_2016}.
}
\label{SI_fig_nsx-as3-poisson-ratio}
\end{figure}

\begin{figure}[htp]
\centering
\includegraphics[width=0.85\columnwidth]{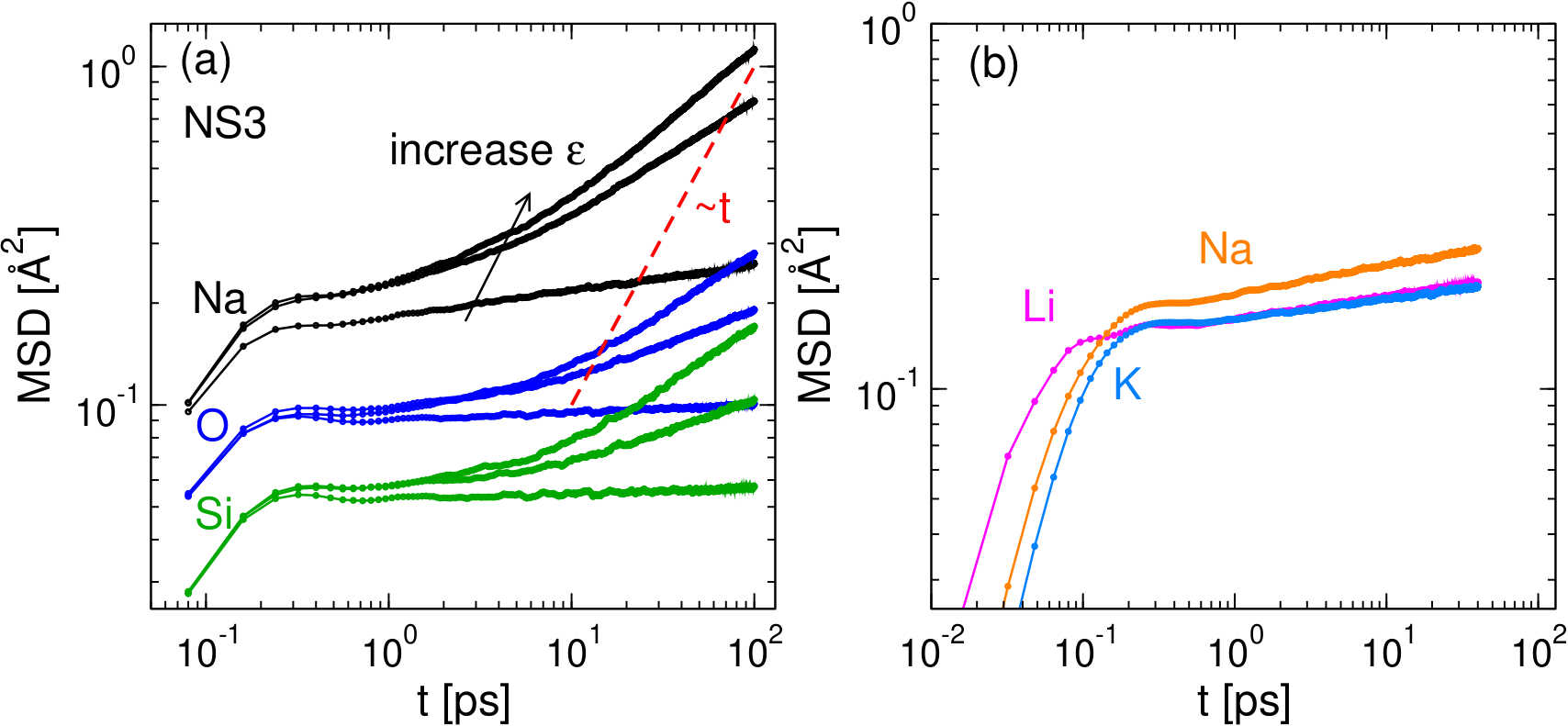}
\caption{ Time dependence of the mean squared displacement (MSD) for different types of atoms. (a) MSD for the three atomic species of NS3. The arrow indicates the increase of strain from 0, 0.1, and 0.2. (b) MSD for the three alkali species of the AS3 glasses for $\varepsilon=0$. The dashed line in (a) has slope 1.0, the value for normal diffusion. The atoms are caged at short times but then become sub-diffusive at long times. Sub-diffusion of Na at long time might be due to stress-activated jumping events, i.e.~an Na atom jumps from one neighborhood site to a neighboring one~\cite{cormack_alkali_2002}. This results in the increase of the atomic displacement but the modifiers are still sub-diffusive. To measure the size of the cage we take the mean of the atomic displacement in the time interval of 0.4-0.5 ps, during which the MSD is still inside the plateau regime. }
\label{SI_fig_msd}
\end{figure}

Figure~\ref{SI_fig_ns3-bend-mix-pot}(a) shows that, in comparison with the behavior predicted by the SHIK potential, the glass put under tension using the reaxFF potential can be deformed much stronger before it fails, irrespective of the potential used for generating the glass sample (i.e.~HO or SHIK). This result is compatible with the findings of our previous study, in which we have shown that the interaction potential has a pronounced influence on the mechanical properties~\cite{zhang_potential_2020}. We also mention that in general the SHIK potential outperforms the HO and the reaxFF potentials in reproducing the experimentally measured mechanical properties of the glasses~\cite{lower_inert_2004,kurkjian_intrinsic_2001}. 

Fig.~\ref{SI_fig_ns3-bend-mix-pot}(b) presents the ratio between stress and strain, $\sigma/\varepsilon$, as a function of $\varepsilon$. Since for these simulations the system size were relatively small, the resulting stress-strain curves have a larger fluctuations than the ones in the main text and hence the calculation of their derivatives is not very reliable. Therefore we look here at the ratio $\sigma/\varepsilon$, which carries a similar information as the derivative. One recognizes that both the SHIK and reaxFF potentials predict a bend in the curve, marked by the dashed line, irrespective of the initial glass structure. This bend in $\sigma/\epsilon$ corresponds to the bend in the tangent modulus of the NS3 glass and one recognizes that its location depends on the interaction potential: The reaxFF potential predicts the bend at larger strains than the SHIK potential.  The fact that the reaxFF potential as well as the SHIK potential predict qualitatively the same anomalous behavior in the stiffness of NS3 glass is strong evidence that this anomaly is indeed a feature of alkali-rich glasses. We also note that the HO potential does not predict the same anomalous behavior of the tangent modulus. It can be speculated that the difference between these curves in Fig.~\ref{SI_fig_ns3-bend-mix-pot} is related to the way the interaction potential was developed, i.e.~what functional form and reference data were considered for describing the interaction between atoms.
\clearpage

\noindent{\bf 2. Supplementary figures}

\end{document}